\documentstyle[prl,aps,epsf,multicol]{revtex}
\begin{document}


\bibliographystyle{prsty}

\title{Single Electron Tunneling at Large Conductance: 
        The Semiclassical Approach}
\author{Georg G\"oppert and Hermann Grabert}
\address{Fakult\"at f\"ur Physik, Albert--Ludwigs--Universit{\"a}t, \\
Hermann--Herder--Stra{\ss}e~3, D--79104 Freiburg, Germany}

\date{\today}
\maketitle
\widetext

\begin{abstract}
We study the linear conductance of single electron devices showing
Coulomb blockade phenomena. Our approach is based on
a formally exact path integral representation describing
electron tunneling nonperturbatively. The
electromagnetic environment of the device is treated in terms of the 
Caldeira-Leggett model. We obtain
the linear conductance from the Kubo formula leading to a
formally exact expression which is evaluated in the semiclassical
limit. Specifically we consider three models. First, the influence 
of an electromagnetic environment of arbitrary impedance on a single
tunnel junction is studied focusing on the limits of large 
tunneling conductance and high to moderately low temperatures. 
The predictions are 
compared with recent experimental data. Second, the conductance of an 
array of $N$ tunnel junctions is determined
in dependence on the length $N$ of the array and the environmental 
impedance. Finally, we consider a single electron
transistor and compare our results for large tunneling conductance 
with experimental findings.
\end{abstract}


\raggedcolumns
\begin{multicols}{2}
\narrowtext

\section{Introduction}

Tunneling of electrons in nanostructures is strongly
affected by Coulomb repulsion. In systems containing metallic tunnel
junctions the interaction can be described by the charging energy
\cite{Nato92} $E_C=e^2/2C$ expressed in terms of a geometrical
capacitance $C$. For weak tunneling and temperatures well below
$E_C/k_B$, tunneling is suppressed by the Coulomb blockade effect. 
This regime is well explored experimentally
\cite{FultonPRL87,DelsingFrequPRL89,KastnerRMP92,PekolaSJPRL96}, and
the phenomena
observed can be explained theoretically 
\cite{GrabertEMPRL90,Averin91,SchoenEMPRB91,IngoldNato92,NazarovTravPRB91} 
by means of perturbation 
theory in the tunneling strength which is 
characterized by
the classical high temperature tunneling conductance $G_T$. This approach
breaks down for conductances comparable to or even larger than 
the conductance quantum
$G_K=e^2/h$. 
When using these devices, e.g.\ as highly sensitive electrometers
\cite{LafargeZPB91}, in detectors \cite{EsteveDetAPL92}, or for 
thermometry
\cite{PekolaThermoPRL94}, a large current signal is desirable meaning 
large
tunneling conductance.
However, higher order processes such as
cotunneling \cite{AverinNato92,KoenigCOTPRL97} lead to a smearing of
Coulomb blockade phenomena and a compromise must be found in 
practice. 
While the strong tunneling regime has been explored
extensively by recent experiments 
\cite{JoyezSETPRL97,JoyezSJPRL98,ToppariSJEPL98,KuzminSETPRB99},
theoretical predictions remain limited. The theoretical work
roughly splits into two groups. On the one hand, higher
order perturbative results
\cite{GrabertBOXPRB94,GrabertBOXPhysica94,GeorgBOXPRL98,KoenigRENPRB98,GeorgChannelEPL99} 
were successful in
explaining some of the recent experimental data, yet, the analysis
typically is restricted to
conductances at most of order $G_K$. Based on the diagrammatic expansion, 
partial resummation
techniques were used to obtain nonperturbative results 
\cite{MatveevBOXJETP91,Schoeller2StatePRB94,FalciScalePRL95,ZaikinNCAPRB94},
however, for a restricted set of charge states. The arbitrary cutoff
necessary in these latter theories limits their use for 
direct comparison with
experimental findings. Further progress can be made 
by using
perturbative renormalization group techniques 
\cite{KoenigRGPRL98,PohjolaDoublePRB99}.
Apart from these approaches based on diagrams generated by treating
tunneling as a perturbation, a formally exact path integral
expression \cite{SchoenREP90} including all orders in the tunneling
conductance may serve as a starting point for analytical
predictions 
\cite{BenJacPRL83,ZaikinECPRL91,ZaikinSJPRB92,ZaikinSETJETP96,WangBOXPRB96} 
and numerical calculations 
\cite{WangMCEPL97,ZwergerBOXPRL97,HerreroMCPRB99}. 
While perturbation theory in the tunneling term usually starts from
states with definite electric charge, this latter approach employs the
canonically conjugate phase variable and thus is well adapted to
situations where the charge is smeared by thermal or quantum
fluctuations.
In this work we use the path integral
approach to determine the linear conductance of single
electron devices in the semiclassical limit. Some limiting cases of
the results presented here were published in short form previously
\cite{GeorgSJPRB97,GeorgSETPRB98,GeorgSEMICRAS99}. 
Here we give a fuller account of the
approach and apply it to a larger variety of systems.

The paper is organized as follows: In Sec.~{\rm II} we introduce
the Hamiltonians of a tunnel junction and of the electromagnetic
environment, respectively. We then explain the general method 
of calculating 
the linear conductance from the Kubo formula with the help of a 
generating functional. In Sec.~{\rm III} the case of a single tunnel
junction embedded in an electromagnetic environment of arbitrary
impedance is considered. We use this example to derive the 
effective action characterizing the generating functional which is
employed also in 
subsequent sections with 
adequate generalizations. We evaluate
the conductance in the semiclassical limit appropriate for high
temperatures and/or large tunneling conductance and compare the 
results with experimental findings by Joyez {\it et al}.\ 
\cite{JoyezSJPRL98} and by Farhangfar 
{\it et al}.\ \cite{ToppariSJEPL98}. 
As a first extension of the 
method, we consider in Sec.~{\rm IV} a linear array of tunnel 
junctions embedded 
in an electromagnetic environment. The conductance of the array 
is determined in the high temperature limit. Specifically, we 
study the effect of the environmental impedance on the conductance and
show that with increasing length of the array the influence of the
environment is strongly suppressed. In Sec.~{\rm V} we turn to a 
single electron transistor (SET). Here, we go beyond
leading order in the semiclassical expansion and determine the 
conductance in dependence on the gate voltage. The findings are
compared with experimental data by Joyez {\it et al}.\ \cite{JoyezSETPRL97}
for SETs in the strong tunneling
regime. We conclude and discuss possible
extensions in Sec.~{\rm VI}.

\section{Model and General Method}

In this section we introduce the Hamiltonian for a
single tunnel junction and model the electromagnetic environment in 
terms of a set of LC circuits. A metal - oxide layer - metal
tunnel junction consists of two metallic leads 
separated by a thin
oxide layer \cite{Nato92,DolanLithoPhysica88}. 
Provided the screening length in the metal is small compared to
typical electrode and oxide barrier dimensions, one may introduce 
a geometrical
capacitance $C$. The energy shift for an electron tunneling from
one lead to the other is determined by the charging energy $E_C=e^2/2C$.
The corresponding Coulomb Hamiltonian reads
\begin{equation}
 H_{\rm C}(Q) = \frac{Q^2}{2C},
\label{eq:HamCou}
\end{equation}
where $Q$ is the charge operator on the capacitance. The leads 
are described in second quantization by
\begin{equation}
 H_{\rm qp}
 =
 \sum_{k\sigma} \epsilon_{k\sigma} a^\dagger_{k\sigma} a_{k\sigma} +
 \sum_{q\sigma} \epsilon_{q\sigma} a^\dagger_{q\sigma} a_{q\sigma} ,
\label{eq:HamQP}
\end{equation}
where the $\epsilon_{p\sigma}$ are quasiparticle energies,
and $a^\dagger_{p\sigma}$ and $a_{p\sigma}$ are creation
and annihilation operators for states on the two
electrodes, respectively. The indices $p=k$, $q$ are
longitudinal wave numbers and $\sigma$ is the channel index including 
transversal and spin quantum numbers.
Provided the tunneling amplitudes are small, we may describe 
barrier transmission by a tunneling Hamiltonian 
\cite{Nato92,BardeenTunnelPRL61}
\begin{equation}
 H_{\rm T}(\varphi)
 =
 \sum_{kq\sigma}
 \left(
  t_{kq\sigma} a^\dagger_{k\sigma} a_{q\sigma} \Lambda
  + {\rm H.c.}
 \right) ,
\label{eq:HamTun}
\end{equation}
preserving the channel index $\sigma$. Here
$t_{kq\sigma}$ is the tunneling amplitude and $\Lambda$ the charge
shift operator obeying $\Lambda^\dagger Q \Lambda = Q+e$. Defining
a conjugate phase $\varphi$ by $[Q,\varphi] = ie$, we may write
\begin{equation}
 \Lambda= \exp(-i\varphi).
\label{eq:ShiftOp}
\end{equation}
The total Hamiltonian of a tunnel junction then reads
\begin{equation}
 H_{\rm J}(Q,\varphi)
 =
 H_{\rm C}(Q)+H_{\rm qp}+H_{\rm T}(\varphi) ,
\label{eq:HamTJ}
\end{equation}
where the dependence on the charge and
conjugate phase operators is made explicit to emphasize the
similarity between the charging energy and a kinetic energy and 
between the tunneling
Hamiltonian and an effective potential energy.

The electromagnetic environment can be described by a Caldeira-Leggett 
model \cite{CaldeiraAP83} as a linear combination of LC circuits
\begin{equation}
 H_{\rm em}(\varphi)
 =
 \sum_{n=1}^N
 \left[
   \frac{Q_n^2}{2C_n} +
   \frac{1}{2L_n}  \left(\frac{\hbar}{e}\right)^2
   \left( \varphi - \phi_n  \right)^2
 \right] ,
\label{eq:HamEM}
\end{equation}
coupled to the phase operator $\varphi$ of the device. 
The parameters of the LC-circuits are related to the
environmental admittance by
\begin{equation}
 Y(\omega)
 = 
 \sum_{n=1}^N \frac{\pi}{L_n} 
 \left[
   \delta(\omega + \omega_n) + \delta(\omega - \omega_n)
 \right] ,
\label{eq:admittance}
\end{equation}
where the $\omega_n=1/\sqrt{L_n C_n}$ are the eigenfrequencies 
of the oscillators. 
A single electron tunneling device consists of tunnel
junctions, capacitances, and admittances. The Hamiltonian of this
system can be constituted from the elements discussed above. 
Then, the bosonic 
degrees of freedom of the admittance and the fermionic degrees of
freedom of the electrodes may be traced out. In the next section we
exemplarily derive the Hamiltonian and the effective action for a
tunnel junction embedded in an environment of arbitrary impedance.

Specifically, in this article we determine the linear conductance 
$G=\partial I/ \partial V |_{V=0}$ of single electron devices.
Here $I$ is the measured current and $V$ the applied voltage.
Within linear response theory one may use the Kubo formula for the
linear conductance
\begin{equation}
 G(\omega)
 =
   \frac{1}{i \hbar \omega }
    \lim_{i \nu_n \rightarrow \omega  + i \delta}
    \int^{\hbar \beta}_0 \! d \tau\, e^{i \nu_n \tau}
    \langle I^{(1)}( \tau) I^{(2)}(0)\rangle  \; ,
\label{eq:Kubo}
\end{equation}
where $I^{(1)}$ is the measured current and $I^{(2)}$ a
current operator determined by the coupling to the applied voltage
$V$, see below. The $\nu_n=2 \pi n /\hbar \beta$ are Matsubara frequencies.
Correlation functions can be written as variational
derivatives
\begin{equation}
 \langle I^{(1)}( \tau) I^{(2)}(\tau')\rangle
 =
 \left.
 \frac{\hbar^2}{Z[0,0]}
 \frac{\delta^2 Z[\xi_1,\xi_2]}
      {\delta \xi_1(\tau) \delta \xi_2(\tau')}
 \right|_{\xi_i \equiv 0}
\label{eq:varder}
\end{equation}
of a generating functional~\cite{BenJacPRL83}
\begin{equation}
 Z[\xi_1,\xi_2]
 =
 {\rm tr} \, T_\tau \exp \!\!
 \left\{
  \!-\frac{1}{\hbar}\int_0^{\hbar \beta} \!\!\! d\tau
  \bigg[
    H -\! \sum_{i=1,2} \! I^{(i)}\xi_i(\tau)
  \bigg] \!
 \right\}
\label{eq:genfunc}
\end{equation}
depending on auxiliary fields $\xi_i$. Here, the Hamiltonian $H$
describes the system at vanishing external voltage, $V=0$, and
$T_\tau$ is the Matsubara time ordering operator. In subsequent
sections we apply this generating functional approach to derive an
explicit expression for the linear conductance in the semiclassical
regime.

\section{Tunnel Junction with Environment}

\subsection{Generating Functional}

\begin{figure}
\begin{center}
\leavevmode
\epsfxsize=0.35 \textwidth
\epsfbox{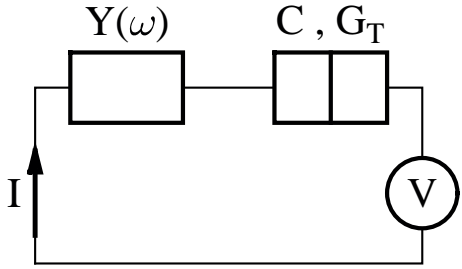}
\end{center}
\caption{Circuit diagram of a tunnel junction in series with an admittance.}
\label{fig:fig1}
\end{figure}

We consider a tunnel junction characterized by the geometrical
capacitance $C$ and the tunneling
conductance $G_T$ embedded in an electromagnetic environment. Via
network transformations it is always possible to transform the
environmental degrees of freedom into an admittance $Y(\omega)$ in series
with the junction biased by a voltage source $V$, {\it cf}.\ 
Fig.~\ref{fig:fig1}. 
In this subsection we obtain the effective action
characterizing the generating functional introduced above. Readers
familiar with path integral techniques for single electron devices may
directly proceed to the next subsection.

The Hamiltonian is given by 
$H=H_J(Q_J,\varphi_J) + H_{\rm em}(\varphi_{\rm em})$. Here the phases
$\varphi_J$ and $\varphi_{\rm em}$ are related to the voltages $V_J$
and $V_{\rm em}$ across the tunnel junction and the admittance,
respectively, by $\dot{\varphi}_J=\scriptsize{\frac{e}{\hbar}}V_J$ and
$\dot{\varphi}_{\rm em}=\scriptsize{\frac{e}{\hbar}}V_{\rm em}$. 
Further, one has to
take care of constraints for the variables imposed by the circuit.
Using Kirchhoff's law for the voltages, we find that the sum of the
phases in the circuit loop in Fig.~\ref{fig:fig1} 
has to be constant, i.e.\
$\varphi_J+\varphi_{\rm em}+\psi=const.$, where we have described the
voltage source in terms of an additional phase \cite{IngoldNato92}
\begin{equation}
 \psi(t)
 =
 \frac{e}{\hbar} \int_{-\infty}^t dt' V(t')  .
\label{eq:PhasePsi}
\end{equation}
Similar relations hold for each loop of more complicated circuits.
For an adequate handling of these constraints we start from the 
Lagrangian description, ${\cal L}=T-U$. In general, the kinetic energy
$T$ is given by the sum of Coulomb energy terms and the effective 
potentials are 
the tunneling and environmental Hamiltonians. The constraints are
naturally implemented by expressing the variables through generalized
coordinates. Defining generalized momenta in the standard way, 
one can derive the 
Hamiltonian via a Legendre transformation.
To define conjugate momenta non-ambiguously, we use the discrete
Caldeira-Leggett model and perform the continuum limit only
afterwards. Shunt capacitors need to be treated separately and will be
discussed in Sec.~{\rm V}. 
Since $\psi(t)$ is
controlled externally, the phase $\varphi_{\rm em}$ may be eliminated
in favor of $\varphi_J \equiv \varphi$ and we may write
\begin{equation}
 H_{\rm JE}(Q,\varphi)
 =
 H_{\rm J}(Q,\varphi) + H_{\rm em}(\varphi + \psi) ,
\label{eq:HamSJ}
\end{equation}
where 
$Q=\scriptsize{\frac{e}{\hbar}}\partial {\cal L}/\partial \dot{\varphi}$
is the momentum canonically conjugate to $\varphi$. 
In the second term, we have absorbed the minus sign in front of 
$\varphi + \psi$ into the
arbitrary definition of the sign of the phase of the environment. 
The current may be defined as the time derivative of the charge
\begin{equation}
 \dot{Q}
 = 
 \frac{i}{\hbar} [H_{\rm JE},Q]
 =
 I_T + I_{\rm em} ,
\label{eq:GenCurrent}
\end{equation}
where 
\begin{equation}
 I_T
 =
 \frac{i}{\hbar} [H_{\rm T}(\varphi),Q]
 =
 - \frac{i e}{\hbar} \sum_{kq\sigma}
 \left(
  t_{kq\sigma} a^\dagger_{k\sigma} a_{q\sigma} \Lambda
      - {\rm H.c.}
 \right) 
\label{eq:TunCurrent}
\end{equation}
is the current through the tunnel junction and 
\begin{equation}
 I_{\rm em}(\varphi)
 =
 \frac{i}{\hbar} [H_{\rm em}(\varphi),Q]
 =
 \frac{\hbar}{e} \sum_{n=1}^N \frac{1}{L_n}(\varphi - \phi_n) 
\label{eq:IROperator}
\end{equation}
the current through the admittance
at vanishing external voltage. To determine
the linear conductance $(\ref{eq:Kubo})$, we choose the measured
current $I^{(1)}$
to be the current $I_{\rm em}$, and $I^{(2)}$ 
follows from 
the coupling to the phase $\psi$ in linear approximation:
$H_{\rm JE}(Q,\varphi,\delta \psi) = H_{\rm JE}(Q,\varphi) + 
\scriptsize{\frac{\hbar}{e}} I^{(2)} \delta \psi$. 
Via a unitary transformation $U=\exp(-i \kappa_1 \psi Q/e)$, it is
always possible to write the external voltage partly as a shift of 
the phase variable $\varphi$. Using 
$U^\dagger \varphi U=\varphi-\kappa_1 \psi$ and the general relation
$H' = U^\dagger H U + i\hbar U^\dagger
\scriptsize{\frac{\partial}{\partial t}}U$,
we get
\begin{equation} 
 H_{\rm JE}'(Q,\varphi)
 =
 H_{\rm J}(Q+\kappa_1 V C,\varphi - \kappa_1 \psi) + 
 H_{\rm em}(\varphi + \kappa_2 \psi) ,
\label{eq:HamSJtrafo}
\end{equation}
where $\kappa_1$ is an arbitrary shift and $\kappa_2=1-\kappa_1$.
Here we choose $\kappa_1=0$ so that the voltage couples solely to
the environmental degrees of freedom and then get 
$H_{\rm em}(\varphi +\delta \psi) = H_{\rm em}(\varphi) + 
\scriptsize{\frac{\hbar}{e}} I_{\rm em} \delta \psi$.
Hence, in this case $I^{(2)}$ coincides with
the measured current $I^{(2)}=I^{(1)}=I_{\rm em}$. To derive
the path integral representation of the generating functional
$(\ref{eq:genfunc})$, we define
\begin{equation}
 \widetilde{H}_{\rm em}(\varphi)  
 \! = \!
 H_{\rm em}(\varphi) - \xi(\tau) I_{\rm em}(\varphi)
 \! = \!
 H_{\rm em}[\varphi - \frac{e}{\hbar} \xi(\tau)] + {\rm ind.}
\label{eq:HamTildeEM}
\end{equation}
where {\rm ind.} denotes a $\varphi$-independent term that may be
omitted. Further, we separate the exponential in
Eq.~$(\ref{eq:genfunc})$ into a free part $A_0(\hbar \beta)$ for 
vanishing 
tunneling and a tunneling part $A_T(\hbar \beta)$ according to
\begin{equation}
 T_\tau
 e^{-\frac{1}{\hbar} \int_0^{\hbar \beta} d\tau
 \left( \widetilde{H}_0 + H_T \right)}
 =
 A_0(\hbar \beta) A_T(\hbar \beta)   ,
\label{eq:intpic}
\end{equation}
where
\begin{equation}
 A_0(\tau)
 =
 T_\tau
 e^{
  -\frac{1}{\hbar} \int_0^{\tau} d\tau'  \widetilde{H}_0(\tau')
    }
\label{eq:intprop}
\end{equation}
describes the system in presence of the unperturbed 
Hamiltonian 
$\widetilde{H}_0 = \widetilde{H}_{\rm em}+H_{\rm C}+H_{\rm qp}$. 
Using the series expansion of $A_T(\tau)$ in powers of $H_T$ and
separating the trace in Eq.~$(\ref{eq:genfunc})$ 
into partial traces over the charge degrees of freedom of the device, 
the quasiparticle components, and the environmental degrees of freedom,
we obtain an expression of the generating functional as a sum of  
averages of the unperturbed system, {\it cf}.\ \cite{GrabertBOXPRB94}.
Due to the Coulomb interaction Hamiltonian $H_C$ in the unperturbed 
Hamiltonian $\widetilde{H}_0$, contributions of a given 
order in $H_T$ cannot simply be
evaluated with the help of Wick's theorem, however,
the partial traces over the quasiparticle components are averages
weighted with the free fermionic density matrix 
$\sim \exp(-\beta H_{qp})$ and, accordingly, 
products of quasi-particle creation and annihilation operators in the
interaction picture decompose into
products of two-pair correlators. In the limit of
large channel number ${\cal N}=\sum_\sigma 1 \gg 1$, only specific
combinations of contractions contribute
that may be written in terms of
two-time correlators of the tunneling Hamiltonian
%
\begin{eqnarray}
 && \!\!\!\!\!\! G(\tau, \tau') 
 =
 \frac{1}{\hbar^2}
 \langle H_{\rm T}(\tau)H_{\rm T}(\tau') \rangle_{\rm qp}
\nonumber             \\
 &=&
 \frac{t^2}{\hbar^2}
 \sum_{k_1 q_1 \sigma_1}
 \sum_{k_2 q_2 \sigma_2}
 \sum_{\zeta_1,\zeta_2 = \pm}
   \zeta_1 \zeta_2 \Lambda^{\zeta_1}(\tau) \Lambda^{-\zeta_2}(\tau')
\nonumber             \\
 &&
 \langle
    a^{\zeta_1}_{k_1\sigma_1}(\tau) a^{-\zeta_2}_{k_2\sigma_2}(\tau')
 \rangle_{\rm qp}
 \langle
    a^{-\zeta_1}_{q_1\sigma_1}(\tau) a^{\zeta_2}_{q_2\sigma_2}(\tau')
 \rangle_{\rm qp}
   \nonumber              \\
 &=&
 \frac{t^2}{\hbar^2}
 \sum_{k q \sigma  \zeta}
   \Lambda^{\zeta}(\tau) \Lambda^{-\zeta}(\tau')
 \frac{
  e^{\zeta(\tau'-\tau)(\epsilon_{k\sigma}-\epsilon_{q\sigma})}
      }{
    (1+e^{\zeta \beta \epsilon_{k\sigma}})
    (1+e^{-\zeta \beta \epsilon_{q\sigma}})
       }  ,
\label{eq:TunnelTauD}
\end{eqnarray}
%
with a real averaged tunneling matrix
element $t=\overline{t_{k q \sigma}}$.
Here $\langle \ldots \rangle_{\rm qp}$ denotes the thermal average over
the quasiparticles with Hamiltonian $H_{\rm qp}$. The time
dependence in the interaction picture reads
\begin{equation}
 H_T(\tau)
 =\exp\left( \frac{\tau}{\hbar} H_{\rm qp} \right) H_T 
  \exp\left(- \frac{\tau}{\hbar} H_{\rm qp} \right)
\end{equation}
and
\begin{equation}
 \Lambda^\zeta(\tau)
 =
 A_0(-\tau)\Lambda^\zeta A_0(\tau) .
\end{equation} 
Further, we have introduced the notation $a^+ = a^\dagger$,
$a^-=a$, and $\Lambda^{\pm} = \exp(\mp i \varphi)$.
Performing the continuum limit for the longitudinal quantum
numbers $k$ and $q$, we find 
\begin{equation}
 G(\tau,\tau')
 =
 \frac{1}{\hbar}
 \frac{G_T}{G_K} \alpha(\tau-\tau')
 \left[
   \Lambda^\dagger(\tau) \Lambda(\tau') +
   \Lambda^\dagger(\tau') \Lambda(\tau)
 \right]
\label{eq:TunnelTauC}
\end{equation}
where $G_T/G_K= 4\pi^2 t^2 {\cal N} \rho \rho'$ is the classical
dimensionless tunneling conductance with the densities of states 
$\rho$ and $\rho'$ at the Fermi
level in the left and right electrode, respectively. In our approach 
the limit of
strong tunneling is defined by ${\cal N}\gg 1$, $t^2 \rho \rho' \ll 1$
such that $4\pi^2 t^2 {\cal N} \rho \rho' \gg 1$. Since for
lithographically fabricated metallic tunnel junctions typically 
${\cal N} {\tiny{\lower 2pt \hbox{$>$} 
  \atop \raise 5pt \hbox{$\sim$}}} 10^4$,
$G_T/G_K$ can become very large, although
each single channel is weakly transmitting only.

The quasiparticle excitations generated by $H_T$
are described by an electron-hole pair Green function
\cite{GrabertBOXPRB94}
\begin{eqnarray}
 \alpha(\tau)
 &=&
 \frac{1}{4 \pi^2 {\cal N} \hbar \rho \rho'} 
 \sum_{k q \sigma}
  \frac{
    e^{(\epsilon_{k\sigma}-\epsilon_{q\sigma}) \tau}
       }{
    (1+e^{- \beta \epsilon_{k\sigma}})
    (1+e^{\beta \epsilon_{q\sigma}})
       }
   \nonumber          \\
 &=&
 \frac{\hbar}{4\pi^2}
 \int_{-\infty}^{\infty} d\epsilon   
  \frac{
     \epsilon \, e^{-|\epsilon|/D}
       }{
    1-e^{-\hbar \beta \epsilon}
        } e^{-\tau \epsilon}
\label{eq:AlphaTau}
\end{eqnarray}
where the electron and hole propagate on different electrodes. $D$
is the electronic bandwidth which may be set to infinity at the end
of the calculation since $D\gg E_C, k_BT$. 
Due to analytic properties of thermal Green
functions, we may write
\begin{equation}
 \alpha(\tau)
 =
 \frac{1}{\hbar \beta} \sum_{n=-\infty}^\infty
 \widetilde{\alpha}(\nu_n) e^{-i\nu_n \tau}
\label{eq:AlphaFou}
\end{equation}
with Fourier coefficients
\begin{equation}
 \widetilde{\alpha}(\nu_n)
 =
 -\frac{\hbar}{4\pi} |\nu_n| e^{-|\nu_n|/D}.
\label{eq:AlphaCoeff}
\end{equation}
Here and in the remainder of the article the absolute value is defined by
\begin{equation}
 |z| =
  \left\{
   \begin{array}{c@{\qquad}r}
    z  & \mbox{Re}(z)>0   \\
    -z  &  \mbox{Re}(z)<0
   \end{array}
  \right._,
\label{eq:DefAbs}
\end{equation}
which leads to a unique analytical continuation \cite{BaymGreensJMP61} of
the Fourier coefficients $(\ref{eq:AlphaCoeff})$. Along these lines
the partial traces
over the quasiparticle components may be evaluated in terms of the
tunneling kernel $\alpha(\tau)$. 

To proceed we need to consider next
the partial trace over the charge degrees of freedom. It
is convenient, to change to the
phase representation and insert identity operators
$\int d\varphi_\tau |\varphi_\tau \rangle \langle \varphi_\tau |$ 
at each imaginary time slice 
$\tau_n=\scriptsize{\frac{\hbar \beta}{N}}n, n=0 \ldots , N$ 
with $N \rightarrow \infty$.
The charge shift operators in the interaction picture then become
$\Lambda^\pm (\tau)=\exp(\mp i \varphi_\tau)$.
Dividing the generating functional $(\ref{eq:genfunc})$  
by the quasiparticle partition
function ${\rm tr}_{\rm qp}\exp(-\beta H_{\rm qp})$ 
which has no effect on the correlator $(\ref{eq:varder})$, we get
\begin{eqnarray}
 Z_{\rm JE}[\xi]
 &=&
 \int {\cal D}[\varphi] 
 \prod_{n=1}^N \int {\cal D}[\phi_n]   
 \exp
 \left\{
    - \frac{1}{\hbar} \, S_0 [\varphi, \phi_n, \xi]
 \right\} 
     \nonumber    \\
&&
 \sum_{m=0}^\infty \int_0^{\hbar \beta} d\tau_{2m}
 \int_0^{\tau_{2m}} d\tau_{2m-1} \ldots
 \int_0^{\tau_2} d\tau_{1} 
     \nonumber    \\
&&
 \sum_{\rm pairs} \prod_{k=1}^m G(\tau_{k_1},\tau_{k_2}),
\end{eqnarray}
where 
$S_0 [\varphi, \phi_n, \xi]
  = S_C[\varphi] + S_{\rm em}[\varphi, \phi_n, \xi]$ 
contains the environmental and the 
Coulomb actions specified below.
Since the integrand is invariant under exchange of an arbitrary
pair of variables we may extend the integrations to 
$\int_0^{\hbar \beta} d\tau_i$ $(i=1, \ldots , 2m)$ and compensate the
larger integration region by a
factor $1/(2m)!$. Further the sum over pairs leads to a factor 
$(2m-1)!!$. Interchanging integrals and product we get 
\end{multicols} \widetext 
\begin{eqnarray}
 Z_{\rm JE}[\xi]
&=&
 \int {\cal D}[\varphi]
 \prod_{n=1}^N \int {\cal D}[\phi_n]  
 \exp
 \left\{
    - \frac{1}{\hbar} \, S_0 [\varphi, \phi_n, \xi]
 \right\} 
 \sum_{m=0}^\infty \frac{1}{m!}
 \left[
  \frac{1}{2}
  \int_0^{\hbar \beta} d\tau
  \int_0^{\hbar \beta} d\tau'
  G(\tau,\tau')
 \right]^m
     \nonumber    \\
&=&
 \int {\cal D}[\varphi] 
 \prod_{n=1}^N \int {\cal D}[\phi_n]  
 \exp
 \left\{
    - \frac{1}{\hbar} \, S_{\rm JE}[\varphi, \phi_n, \xi]
 \right\} ,
\label{eq:GenFuncSJ1}
\end{eqnarray}
\begin{multicols}{2} \narrowtext
\noindent
where the effective Euclidean action splits into three parts
\begin{equation}
 S_{\rm JE}[\varphi, \phi_n, \xi]
 =
 S_C[\varphi] +S_T[\varphi]  + S_{\rm em}[\varphi, \phi_n, \xi] \, .
\label{eq:ActionSJ1}
\end{equation}
Here
\begin{equation}
 S_C[\varphi]
 =
 \int^{\hbar \beta}_0  d \tau \,
 \frac{\hbar^2C}{2e^2} \, \dot{\varphi}^2
\label{eq:ActionCou}
\end{equation}
describes Coulomb charging and
\begin{equation}
 S_T[\varphi]
 \!=\!
 2 \frac{G_T}{G_K}
 \int^{\hbar \beta}_0 \!\!\! d \tau  \!
 \int^{\hbar \beta}_0 \!\!\! d \tau' 
 \alpha(\tau - \tau')
 \sin^2 \! \left[
   \frac{\varphi(\tau)- \varphi(\tau')}{2}
           \right]
\label{eq:ActionTJ}
\end{equation}
quasi-particle tunneling across the junction. 
The environmental action is given by 
\begin{eqnarray}
 S_{\rm em}[\varphi, \phi_n, \xi] 
 &=&
 \sum_{n=1}^N 
 \int_0^{\hbar \beta} d\tau
 \bigg[
   \frac{\hbar^2 C_n}{2e^2} \dot{\phi}_n^2
     \nonumber              \\
 &&
   +\frac{\hbar^2}{2e^2 L_n}
   \left(
     \varphi-\frac{e}{\hbar}\xi - \phi_n
   \right)^2
 \bigg].
\end{eqnarray}
The remaining trace over environmental degrees of freedom in 
Eq.~$(\ref{eq:GenFuncSJ1})$ can be
evaluated exactly
\cite{GrabertREP88} leading to a quadratic nonlocal action
\begin{eqnarray}
 S_Y[\varphi, \xi]
 &=&
 \frac{1}{2} \int^{\hbar
 \beta}_0  d \tau  \int^{\hbar \beta}_0  d \tau^\prime
 k(\tau - \tau^\prime)
 \bigg[
   \varphi(\tau)  - \frac{e}{\hbar} \xi(\tau) 
     \nonumber              \\
 &&
  - \varphi(\tau' ) + \frac{e}{\hbar} \xi(\tau')
 \bigg]^2,
\label{eq:ActionEM}
\end{eqnarray}
where the kernel $k(\tau)$ can be written as a Fourier series
$(\ref{eq:AlphaFou})$
with coefficients
\begin{equation}
 \widetilde{k}(\nu_n)
  =
  -\frac{\hbar}{4\pi} \frac{\widehat{Y}(|\nu_n|)}{G_K} |\nu_n|  .
\label{eq:kCoeff}
\end{equation}
Here $\widehat{Y}(s)$ is the Laplace transform of the
environmental response
function $Y(t)$, {\it cf.} Ref.\ \cite{GrabertREP88}. Due to
causality, for $\mbox{Re}(s)>0$, one may write
$\widehat{Y}(s)=Y(is)$ where
$Y(\omega)$ is the frequency dependent admittance 
$(\ref{eq:admittance})$ of the environment.

This way the generating functional reads
\begin{equation}
  Z_{\rm JE}[\xi]
 =
 \int {\cal D}[\varphi]  
 \exp
 \left\{
    - \frac{1}{\hbar} \, S_{\rm JE}[\varphi, \xi]
 \right\} ,
\label{eq:GenFuncSJ}
\end{equation}
with the effective action
\begin{equation}
 S_{\rm JE}[\varphi, \xi]
 =
 S_C[\varphi] +S_T[\varphi]  + S_Y[\varphi, \xi] \, .
\label{eq:ActionSJ}
\end{equation}
The explicit form of the generating functional serves as a starting
point to calculate the correlator in the next subsection.

\subsection{Conductance}

We now perform the functional
derivatives in Eq.~$(\ref{eq:varder})$ 
explicitly and get for the correlator \cite{GeorgSJPRB97}
\begin{eqnarray}
 \langle I_{\rm em}(\tau)I_{\rm em}(0)\rangle
 &=&
 \frac{1}{Z_{\rm JE}}  \int
 {\cal D}[\varphi]
 \exp \left\{ - \frac{1}{\hbar} S_{\rm JE}[\varphi,0] \right\}
     \nonumber              \\
 &&
 \times 
 \left( 2 \frac{e^2}{\hbar} k(\tau) +
 I_{\rm em} [\varphi, \tau]  I_{\rm em} [\varphi, 0]
 \right),
\label{eq:IICorrSJ}
\end{eqnarray}
where $Z_{\rm JE}=Z_{\rm JE}[0]$ denotes the partition function.
The current functional $I_{\rm em}[\varphi, \tau]$ arising as
variational derivative of the effective action $(\ref{eq:ActionSJ})$ 
reads
\begin{equation}
 I_{\rm em} [\varphi, \tau]
 =
 \frac{2e}{\hbar}
 \int^{\hbar \beta}_0 d \tau^\prime \,
  k(\tau - \tau^\prime)
     \varphi(\tau^\prime) \, .
\label{eq:CurrFuSJ}
\end{equation}
The conductance $(\ref{eq:Kubo})$ now splits into two pieces.
\begin{equation}
 G_{\rm JE}(\omega) 
 =
 G_{\rm JE}^{(1)}(\omega) + G_{\rm JE}^{(2)}(\omega)
\label{eq:CondSplit} 
\end{equation}
where 
\begin{equation}
 G_{\rm JE}^{(1)}(\omega)
 =
 \frac{1}{i \hbar \omega} \frac{2 e^2}{\hbar}
  \widetilde{k}(-i \omega+\delta)
 =
 Y(\omega)
\label{eq:G1OmegaSJ}
\end{equation}
corresponds to the first term in Eq.~$(\ref{eq:IICorrSJ})$, and
\begin{equation}
 G_{\rm JE}^{(2)}(\omega)
 =
 \frac{1}{i \hbar \omega } F_{\rm JE}(-i\omega + \delta)
\label{eq:Gprime}
\end{equation}
with
\begin{equation}
 F_{\rm JE}(\nu_n)
 =
 \frac{1}{Z_{\rm JE}}\int D[\varphi] \exp
 \left\{
  -\frac{1}{\hbar}S_{\rm JE}[\varphi,0]
 \right\}
 F[\varphi,\nu_n]
\label{eq:C2formal}
\end{equation}
to the second term in Eq.~$(\ref{eq:IICorrSJ})$.
The explicit form of the auxiliary functional $F[\varphi,\nu_n]$ 
in terms of the
Fourier components $\widetilde{\varphi}(\nu_m)$ reads
\begin{equation}
 F[\varphi,\nu_n]
 =
 \frac{4 e^2 \beta}{\hbar} \widetilde{k}(\nu_n) \widetilde{\varphi}(\nu_n)
 \sum_{m=-\infty}^{+\infty} \widetilde{k}(\nu_m)
 \widetilde{\varphi}(\nu_m).
\label{eq:CurrFunc}
\end{equation}
So far no approximations have been made and 
Eqs.~$(\ref{eq:CondSplit})-(\ref{eq:CurrFunc})$ give a 
formally exact representation of the linear conductance.
To proceed we evaluate the path integral $(\ref{eq:C2formal})$ 
in the semiclassical limit.

\subsection{Semiclassical Limit}
\begin{figure}
\begin{center}
\leavevmode
\epsfxsize=0.35 \textwidth
\epsfbox{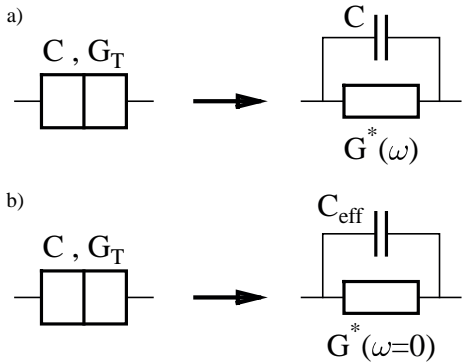}
\end{center}
\caption{Effective circuit diagrams for a tunnel junction in
the semiclassical limit a) for arbitrary frequency and b) in the
low frequency limit.}
\label{fig:fig2}
\end{figure}

The classical trajectory of the phase ${\bar \varphi}$ is defined
by $\delta S_{\rm JE}[{\bar \varphi},0]/\delta {\bar \varphi}=0$,
and we obtain from Eq.~$(\ref{eq:ActionSJ})$ 
${\bar \varphi}=\varphi^0={\rm const}$. 
Since the action is invariant under
a global phase shift, we may put $\varphi^0=0$.
Writing the action in terms of Fourier coefficients of the phase
\begin{equation}
 \widetilde{\varphi}(\nu_n)
 =
 \frac{1}{\hbar \beta} 
 \int_0^{\hbar \beta} d\tau e^{i\nu_n \tau} \varphi(\tau)
\end{equation} 
and expanding in powers of $\widetilde{\varphi}(\nu_n)$, we get 
\begin{equation}
 S_{\rm JE}[\varphi]
 = S_{\rm JE}^0[\varphi] + \sum_{k=2}^\infty S_{\rm JE}^{2k}[\varphi], 
\label{eq:ActionExp}
\end{equation}
where
\begin{equation}
 S_{\rm JE}^0[\varphi]
 =
 \hbar \sum_{n=1}^\infty
   \lambda_{\rm JE} (\nu_n) |\widetilde{\varphi}(\nu_n)|^2
\end{equation}
is the second order variational action with the eigenvalues
\begin{equation}
 \lambda_{\rm JE}(\nu_n)
 =
 \frac{\hbar^2 \beta}{e^2}
 |\nu_n| \left[ \widehat{G}_0(\nu_n) + \widehat{Y}(|\nu_n|) \right].
\label{eq:EigenJE}
\end{equation}
Here
\begin{equation}
 \widehat{G}_0(\nu_n) = |\nu_n| C  + G_T
\label{eq:GTLaplace}
\end{equation}
describes the tunnel junction as a capacitance in parallel with an Ohmic
resistor characterized by the classical tunneling conductance.
Further
\begin{eqnarray}
 S_{\rm JE}^{2k}[\varphi]
 &=&
   \frac{G_T}{G_K}
   \frac{(-1)^{k+1}}{(2k)!} \sum_{l=1}^{2k-1}
   {2k \choose l} (-1)^l \hbar\beta
   \nonumber  \\
 &&  \times  
   {\sum_{n_1, \cdots, n_{2k-1}}}^{\!\!\!\!\!\!\!\! \prime}
   \quad
   \widetilde{\alpha}
   \left(
      -\sum\limits_{p=1}^{l} \nu_{n_p}
   \right)
     \nonumber              \\
 &&  \times
   \widetilde{\varphi}(\nu_{n_1}) 
   \cdots
   \widetilde{\varphi}(\nu_{n_{2k-1}})
   \widetilde{\varphi}
   \left(
     -\sum\limits_{p=1}^{2k-1} \nu_{n_p}
   \right)
 \label{eq:HigherVar} 
\end{eqnarray}
is the variational action of order $2k$.
The summation over the $n_i$ is over all integers with $n_i=0$
omitted. Neglecting sixth and higher 
order terms, we get from Eq.~$(\ref{eq:C2formal})$
\begin{eqnarray}
 F_{\rm JE}(\nu_n)
 &=&
 \frac{4e^2\beta}{\hbar}\,
 \frac{\widetilde k(\nu_n)^2}{\lambda_{\rm JE}(\nu_n)}
 \Bigg[
   1 + \frac{G_T}{G_K} \frac{2\beta}{\lambda_{\rm JE}(\nu_n)}
     \nonumber              \\
 &&  \times
   \sum_{{m=-\infty}\atop {m\ne 0}}^{\infty}
   \frac{\widetilde\alpha(\nu_{n+m})-\widetilde\alpha(\nu_n)
          -\widetilde\alpha(\nu_m)}{\lambda_{\rm JE}(\nu_m)}
 \Bigg]  .
 \label{eq:C_2}
\end{eqnarray}
The convergence of this expansion depends crucially on the eigenvalues
$(\ref{eq:EigenJE})$. To estimate the range of validity of the 
truncated series, we write the
smallest eigenvalue in more appropriate units as
\begin{equation}
 \lambda_{\rm JE}(\nu_1)
 =
  \frac{2 \pi^2}{\beta E_C} + \frac{G_T+\widehat{Y}(\nu_1)}{G_K} .
\label{eq:Eigen1}
\end{equation}
This eigenvalue has to be large compared to $1$,
and we see that the expansion is useful for large conductances
$G_T + \widehat{Y}(\nu_1) \gg G_K$ and/or high temperatures $\beta
E_C \ll 2 \pi^2$. Hence, we effectively expand in powers of 
\begin{equation}
 \epsilon
 =
 \mbox{Min}\left(
   \frac{G_K}{G_T+\widehat{Y}(\nu_1)},
   \frac{\beta E_C}{2\pi^2}
 \right).
\end{equation}
Performing the limit 
$i\nu_n \rightarrow \omega +i\delta$, the analytically continued 
eigenvalue $(\ref{eq:EigenJE})$ reads
\begin{equation}
 \lambda_{\rm JE}(-i \omega)
 =
 -i \omega \frac{\hbar^2 \beta}{e^2}
 \left[
 G_0(\omega) + Y(\omega)
 \right],
\label{eq:lambdacont}
\end{equation}
where
\begin{equation}
 G_0(\omega)
 =
 G_T -i\omega C
\label{eq:GTFourier}
\end{equation}
is the analytic continuation of the Laplace transform of Eq.\
$(\ref{eq:GTLaplace})$. The relative minus sign of the capacitive
term is due to the usual definition of the Fourier transform in
quantum mechanics, the electro-technical convention is obtained by
replacing $\omega \rightarrow -\omega$. For small frequencies the
analytically continued eigenvalue $(\ref{eq:lambdacont})$ is no 
longer large
compared to $1$ and we are faced with a problem of order reduction. 
In the limit $i\nu_n \rightarrow \omega + i\delta$ 
each 
$1/\lambda_{\rm JE}(\nu_n)$ 
in Eq.~$(\ref{eq:C_2})$ becomes of order $1$ while the 
$1/\lambda_{\rm JE}(\nu_m)$  
factors for $m\not= n$ 
are not analytically continued and remain of order
$\epsilon$. The correction term of order $\epsilon^2$ in 
Eq.~$(\ref{eq:C_2})$, that is the term proportional to $G_T/G_K$,
becomes of order $\epsilon$ after analytic continuation. Hence we
loose one factor of $\epsilon$. Generally, one finds that the higher
order variational actions $(\ref{eq:HigherVar})$ include at most one 
$1/\lambda_{\rm JE}(\nu_n)$ factor and consequently are reduced 
at most by one order in $\epsilon$. Thus, products of the form
\begin{equation}
 S_{\rm JE}^{2k_1}[\varphi] S_{\rm JE}^{2k_2}[\varphi] 
 \ldots S_{\rm JE}^{2k_l}[\varphi]
\end{equation}
of quartic or higher-order variational actions, as they arise from an
expansion in powers of $\widetilde{\varphi}(\nu_k)$, give quantum 
corrections of order
$\epsilon^{k_1+k_2+\ldots+k_l}$ and after analytical continuation of
order 
$\epsilon^{k_1+k_2+\ldots+k_l-l}$ and of
higher orders. This proves that the terms of the expansion of 
$F_{\rm JE}$ given explicitely in Eq.~$(\ref{eq:C_2})$ suffice 
to calculate the leading order quantum corrections. After performing 
the analytical continuation we get 
\begin{equation}
 G_{\rm JE}^{(2)}(\omega)
 =
 -\frac{Y(\omega)^2}{G_0(\omega)+Y(\omega)}
 \left[
  1+ \frac{G_T}{G_0(\omega)+Y(\omega)} {\cal U}(\omega)
 \right]
\end{equation}
with the quantum correction factor 
\begin{equation}
 {\cal U}(\omega)
 =
 \frac{2}{i \omega} \sum_{m=1}^\infty \nu_m
 \left[
  \frac{1}{\lambda_{\rm JE}(\nu_m - i \omega)}
  -\frac{1}{\lambda_{\rm JE}(\nu_m)}
 \right]  .
\label{eq:qmsuppSJ}
\end{equation}
Hence, for the total conductance we may write
\begin{equation}
 G_{\rm JE}(\omega)
 =
 \frac{G_{\rm eff}(\omega) Y(\omega)}
      {G_{\rm eff}(\omega) + Y(\omega)}
\label{eq:GallgJE}
\end{equation}
with an effective linear conductance of the junction
\begin{equation}
 G_{\rm eff}(\omega)
 =
 G_T
 \left[
   1 - {\cal U}(\omega)
 \right]
   -i \omega C  .
\label{eq:EffLinSJ}
\end{equation}
This describes a linear element 
$G^*(\omega)=G_T[1-{\cal U}(\omega)]$, depending on the whole 
circuit, in parallel with the geometrical
junction capacitance $C$, {\it cf}.\ Fig.~\ref{fig:fig2}a. 
The general form $(\ref{eq:GallgJE})$ is valid only
to first order in $\epsilon$.
A systematic treatment of higher order contributions does
not allow for a description of the tunnel junction in terms of
an effective linear element.
However, a partial resummation of higher order terms according to
a self-consistent harmonic approximation \cite{JoyezSJPRL98,GeorgSCHA99}
leads again to the form $(\ref{eq:GallgJE})$.

\subsection{Results and Comparison with Experimental Data}
\begin{figure}
\begin{center}
\leavevmode
\epsfxsize=0.45 \textwidth
\epsfbox{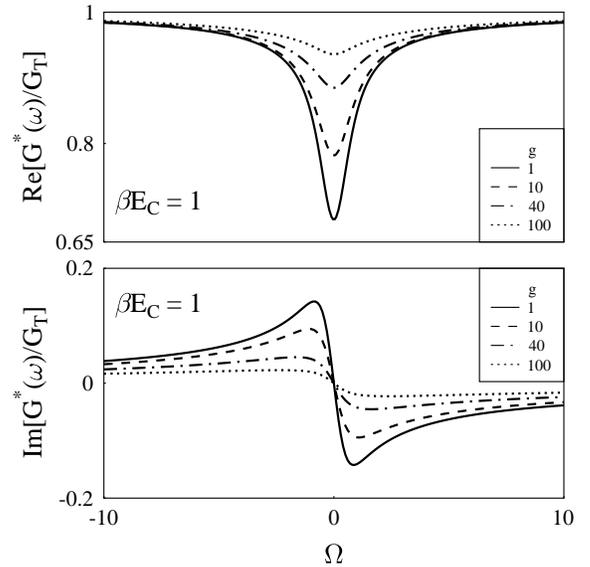}
\end{center}
\caption{Real and imaginary parts of $G^*(\omega)/G_T$ in the 
ohmic damping case for $\beta E_C=1$ and various values of the 
dimensionless conductance
$g$ in dependence on the dimensionless frequency
$\Omega = \hbar \omega /2 \pi E_C$.}
\label{fig:fig3}
\end{figure}

For further discussion and comparison with
experimental data we restrict ourselves to ohmic dissipation
$Y(\omega)=Y$. The effective linear element $(\ref{eq:EffLinSJ})$
then reads
\begin{eqnarray}
 \frac{G^*(\omega)}{G_T}
 &=&
 1-
 \bigg[
  \frac{\psi(1+u+\widetilde{\omega})
        -\psi(1+\widetilde{\omega})}{u}
     \nonumber              \\
 && \qquad
  +
  \frac{\psi(1+u+\widetilde{\omega})
        -\psi(1+u)}{\widetilde{\omega}}
 \bigg] \frac{\beta E_C}{\pi^2},
\label{eq:ohmomegaJE}
\end{eqnarray}
where $\psi$ is the logarithmic derivative of the gamma function and
\begin{equation}
 u
 =
 g \frac{\beta E_C}{2\pi^2}
 , \qquad
 \widetilde{\omega}=\frac{\hbar \beta}{2 \pi i} \omega
\end{equation}
are auxiliary quantities.
We also have introduced the dimensionless parallel conductance
\begin{equation}
 g=(G_T + Y)/G_K.
\end{equation} 
The quantum corrections depend only on this
combination of conductances.
The real and imaginary parts of $G^*(\omega)/G_T$ are depicted in
Fig.~\ref{fig:fig3} for $\beta E_C=1$ and various values of $g$.

The quantum corrections are most pronounced at zero frequency and 
disappear nonalgebraically for
large $\omega$ and/or $u$, due to the logarithmic behavior of the 
psi-function for large arguments.

\begin{figure}
\begin{center}
\leavevmode
\epsfxsize=0.4 \textwidth
\epsfbox{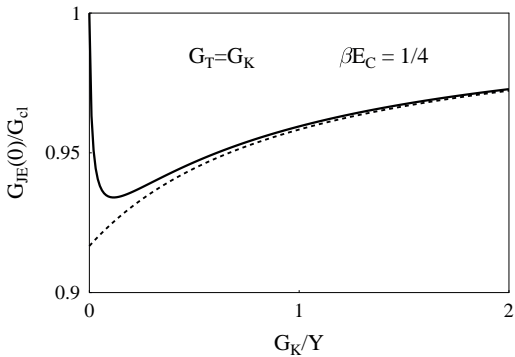}
\end{center}
\caption{The ratio of the total dc conductance $G_{\rm JE}(0)$ and
the classical conductance $G_{\rm cl}=G_T Y / (G_T + Y)$ shown
vs.\ the dimensionless environmental resistance 
$G_K/Y$ for $G_T=G_K$ and $\beta E_C=1/4$. 
The solid line is the result \protect
$(\ref{eq:GallgJE})$ for $\omega=0$ and ohmic damping, 
and the dotted line corresponds to the
approximation $(\ref{eq:entwSJ})$ for moderate to 
small $Y$.} \label{fig:fig4}
\end{figure}

For the dc conductance we get from $(\ref{eq:ohmomegaJE})$
\begin{equation}
 \frac{G^*(\omega = 0)}{G_T}
 =
 1-
 \left[
  \frac{\gamma+\psi(1+u)}{u} + \psi'(1+u)
 \right] \frac{\beta E_C}{\pi^2}
\label{eq:ohm0JE}
\end{equation}
which coincides with our previous result
\cite{GeorgSJPRB97}.
In particular, in the limit of a very low resistance
environment, the total conductance
$(\ref{eq:GallgJE})$ approaches the classical limit nonanalytically, 
{\it cf}.\ Fig.~\ref{fig:fig4},
leading to the asymptotic expansion \cite{GeorgSJPRB97}
\begin{eqnarray}
 &&G_{\rm JE}(\omega = 0)
     \nonumber              \\
 &&
 =
 G_T
 \left[
   1+2\frac{G_K}{Y} \ln\left(\frac{G_K}{Y}\right)
    +{\cal O}\left(\frac{|\ln(\beta E_C)| G_K}{Y}\right)
 \right].
\label{eq:asymptSJ}
\end{eqnarray}
On the other hand for moderate to large environmental resistance,
we may expand Eq.~$(\ref{eq:ohm0JE})$ with respect to $u$ 
leading to a total conductance
\begin{eqnarray}
 && G_{\rm JE}(\omega=0)
     \nonumber              \\
 && 
 = \!
 \frac{G_T Y}{G_T +Y} \!
 \left\{ \!
  1 \! - \! \frac{Y}{G+Y} \frac{\beta E_C}{3}
  \! + \! 
  {\cal O} \left[ (\beta E_C)^2,u\beta E_C \right]
 \! \right\}.
\label{eq:entwSJ}
\end{eqnarray}
This approximation correspond to the dotted line in 
Fig.~\ref{fig:fig4} and remains analytic in the limit of large
environmental conductance where it obviously fails.

\begin{figure}
\begin{center}
\leavevmode
\epsfxsize=0.45 \textwidth
\epsfbox{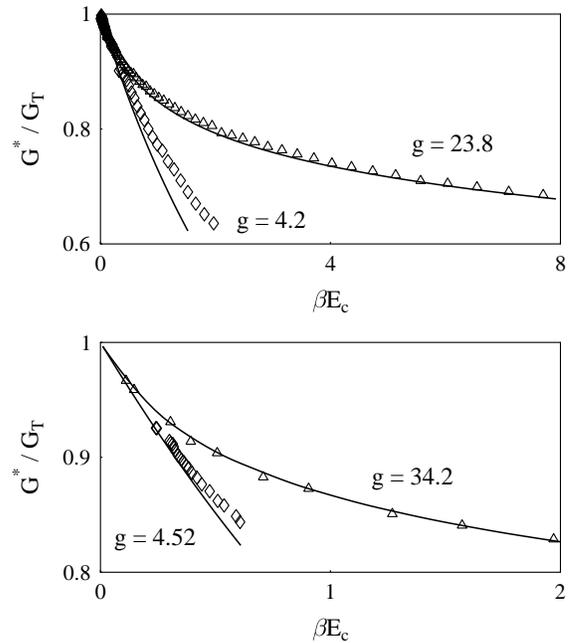}
\end{center}
\caption{The renormalized conductance $(\protect\ref{eq:ohm0JE})$ 
versus the dimensionless
temperature compared with experimental data
by Joyez {\it et al}.\ \protect\cite{JoyezSJPRL98}
for dimensionless parallel
conductance $g=4.2$ and $23.8$ (upper plot)
and with experimental data
by Farhangfar {\it et al}.\ \protect\cite{ToppariSJEPL98}
for $g=4.52$ and $34.2$  
(lower plot).}
\label{fig:fig5}
\end{figure}

In Fig.~\ref{fig:fig5} we compare our prediction 
$(\ref{eq:ohm0JE})$ with recent experimental
data by Joyez {\it et al}.\ \cite{JoyezSJPRL98} for
dimensionless conductance
$g=4.2$ and $23.8$ (upper plot) and by Farhangfar 
{\it et al}.\ \cite{ToppariSJEPL98}
for $g=4.52$ and $g=34.2$ (lower plot). 
Fig.~\ref{fig:fig5} shows
that in the limit of large conductance we are able to explain the whole
range of temperatures explored experimentally, whereas for
moderate conductance only the high
temperature part is covered by the semiclassical theory.
Here, the parameters $g$ and $E_C$ have not been
adjusted to improve the fit but coincide with those given in the
experimental papers.

\begin{figure}
\begin{center}
\leavevmode
\epsfxsize=0.4 \textwidth
\epsfbox{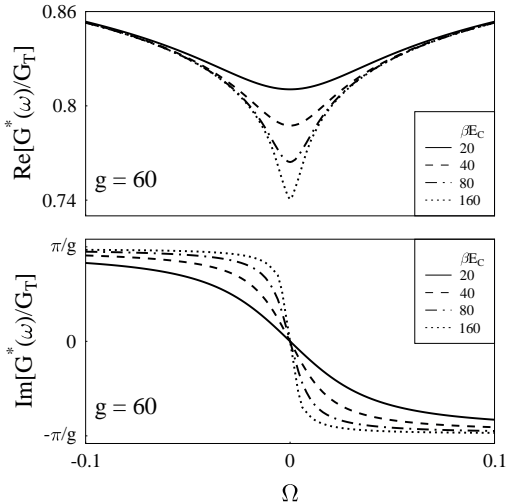}
\end{center}
\caption{Real and imaginary parts of $G^*(\omega)/G_T$ in the 
Ohmic damping
case for dimensionless conductance $g=60$ and inverse temperatures
$\beta E_C=20,40,80$, and $160$ in dependence on the dimensionless
frequency $\Omega = \omega \hbar/2 \pi E_C$.}
\label{fig:fig6}
\end{figure}

We conclude this section with some remarks on the frequency dependence
of the conductance that has not been studied experimentally, so far.
For small frequencies we may expand the result $(\ref{eq:ohmomegaJE})$
and write
\begin{equation}
 G^*(\omega)
 = 
 G^*(\omega = 0) -i \omega C^* + {\cal O}(\omega^2)  ,
\end{equation}
where $C^*$ leads to a renormalization of the junction capacitance $C$.
The renormalized capacitance $C_{\rm eff}=C+C^*$ reads
\begin{equation}
 \frac{C_{\rm eff}}{C}
 \! = \!
 1+ \frac{G_T}{G_K}
 \left[
  \frac{\frac{\pi^2}{3} - 2 \psi'(1+u)}{u}
  - \psi''(1+u)
 \right]
 \frac{(\beta E_C)^2}{4 \pi^4}.
\label{eq:cap0}
\end{equation}
The correction shows a quadratic dependence on
$\beta E_C$ and therefore 
is suppressed at high temperatures. It also vanishes linearly for
large conductance $g$ due to the analytical properties of the psi
function. The semiclassical treatment covers only the region of weak
Coulomb blockade.
Whereas for small tunneling conductance low temperatures imply
strong Coulomb blockade, these effects are suppressed for
highly conducting tunnel junctions and the semiclassical theory is
restored. A closer examination of Eq.~$(\ref{eq:ohmomegaJE})$
shows that for $g \gg 2 \ln(\beta E_C)$ the quantum corrections are
always small. For fixed $g \gg 1$, 
our predictions are therefore valid for a very large
range of temperatures covering in fact the entire range of parameters
presently attainable experimentally for metallic junctions with strong
tunneling
\cite{JoyezSJPRL98,ToppariSJEPL98,KuzminSETPRB99}.
Fig.~\ref{fig:fig6} depicts the real and imaginary parts of
$G^*(\omega)/G_T$ for $g=60$ and various temperatures.
With decreasing temperature 
the real part shows for $\omega=0$ a logarithmic
decrease,
$G^*/G_T=1-2\ln(\beta E_C)/g$, as long as
$k_B T \gg E_C \exp(-g/2)$. Thus, for large conductance the
semiclassical treatment is an effective high temperature expansion
valid for $k_B T$ large compared with the renormalized 
charging energy $E_C^* \approx E_C \exp(-g/2)$ 
\cite{ZaikinECPRL91,WangBOXPRB96}. Moreover, the 
analytical form of the quantum
corrections indicates that Coulomb blockade survives for arbitrary large
conductance but becomes strong only for temperatures below $E_C^*/k_B$.

\begin{figure}
\begin{center}
\leavevmode
\epsfxsize=0.45 \textwidth
\epsfbox{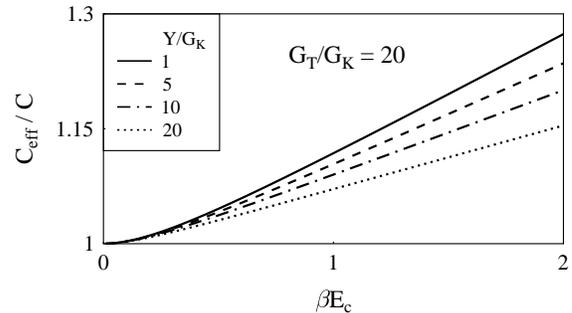}
\end{center}
\caption{Renormalized capacitance $C_{\rm eff}/C$ in the Ohmic 
damping case
for tunneling conductance $G_T/G_K=20$ and various environmental 
conductances
$Y/G_K=1,5,10$, and $20$ in dependence on the dimensionless inverse
temperature $\beta E_C$.}
\label{fig:fig7}
\end{figure}

In the limit $T\rightarrow 0$, $g\rightarrow \infty$ such that
$k_B T \gg E_C \exp(-g/2)$, the imaginary part of $G^*(\omega)$ 
becomes a step function
of width $2 \pi /g$, {\it cf}.\ Fig.~\ref{fig:fig6}, leading to a
divergent renormalized capacitance of the form
$C_{\rm eff}/C= \beta E_C G_T/6(G_T+Y)$. The linear dependence of
$C_{\rm eff}$ on
$\beta$ starts already at very high temperatures,
{\it cf}.\ Fig.~\ref{fig:fig7}, and only saturates for $\beta E_C$ of
order $\exp(g/2)$. 
The large renormalized capacitance is a strong tunneling effect 
due to multiple electron
tunneling and is found likewise for non-Ohmic
environmental impedances. 
In Fig.~\ref{fig:fig7} we show the renormalized capacitance 
$C_{\rm eff}/C$ 
for $G_T/G_K=20$ and various values of $Y/G_K$ in dependence on the 
dimensionless inverse temperature $\beta E_C$. 
Note that the linear behavior of the capacitance starts
already near $\beta E_C = 1$.

The renormalized capacitance describes the frequency dependence of the
conductance for small frequencies $\omega C_{\rm eff} \ll \pi G_T/g$,
{\it cf}.\ Fig.~\ref{fig:fig6}. Rewriting this inequality we get
$\omega \ll 6 k_B T /\hbar \approx 10^{11} T \, {\rm Hz}$, 
where $T$ is the
temperature measured in Kelvin. Thus for all accessible temperatures
the frequency range of strong $1/f$ noise can be avoided, and the effect
predicted should be clearly observable experimentally.

\section{Array of Junctions with Environment}

\subsection{Generating Functional and Conductance}
\begin{figure}
\begin{center}
\leavevmode
\epsfxsize=0.4 \textwidth
\epsfbox{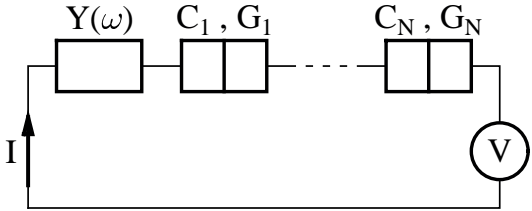}
\end{center}
\caption{Circuit diagram of an array of $N$ tunnel junctions in series
with an admittance $Y(\omega)$.}
\label{fig:fig8}
\end{figure}

As a first extension of the method, we now consider 
linear arrays of $N$ tunnel junctions embedded
in an electromagnetic environment. The junctions
are characterized by classical tunneling conductances $G_j$
and geometrical capacitances $C_j$ in parallel. Like in the previous 
section,
the environment can be transformed into an admittance $Y(\omega)$ in
series with an array of junctions biased by a voltage source
$V$\/, {\it cf}.\ Fig \ref{fig:fig8}. 
We start with a Lagrangian description 
depending on phase variables $\varphi_j$ of each junction
$j=1 \ldots N$ and an environmental phase $\varphi_{\rm em}$ with the 
constraint 
$\sum_{j=1}^N \varphi_j + \varphi_{\rm em} + \psi = const.$,
where $\psi$ describes the applied voltage and is given 
by Eq.\ $(\ref{eq:PhasePsi})$.
Using the $\varphi_j$, $j=1 \ldots N$ as generalized variables we find for
the total Hamiltonian
\begin{equation}
 H_{\rm AE}(\{Q_j\},\{\varphi_j\})
 \! = \!
 \sum_{j=1}^N  H_{\rm J}(Q_j,\varphi_j)
 + H_{\rm em}  \!
 \left( \!
    \sum_{j=1}^N \varphi_j + \psi 
 \! \right),
\end{equation}
with the junction and
environmental Hamiltonians defined by 
Eqs.~$(\ref{eq:HamCou})-(\ref{eq:HamEM})$.
We follow the analysis in the previous section and first derive a 
formally exact
expression for the linear conductance.
As measured current $I^{(1)}$ we choose again the current flowing
through the environmental impedance given by
Eq.~$(\ref{eq:IROperator})$ 
with $\varphi$ replaced by
$\sum_{j=1}^N \varphi_j$ and $Q$ by $\sum_{j=1}^N Q_j$, respectively. 
The second current operator $I^{(2)}$ is determined 
by the linear coupling to
$\psi$ and we get $I^{(1)}=I^{(2)}=I_{\rm em}$. Following the
lines of reasoning in the previous sections, the generating
functional is found to read
\begin{equation}
 Z_{\rm AE}[\xi]
 =
 \int D[\{ \varphi_j \}] \exp
 \left\{
    - \frac{1}{\hbar} \, S_{\rm AE} [\{ \varphi_j \}, \xi]
 \right\} ,
 \label{pathint}
\end{equation}
with the effective Euclidean action
\begin{equation}
 S_{\rm AE}[\{ \varphi_j \},\xi]
 =
 S_Y
 \left[ \sum\nolimits_{i=j}^N \varphi_j , \xi \right] +
 \sum_{i=j}^N
 S_j[\varphi_j]
 \, .
\label{action}
\end{equation}
Here $S_Y$ was introduced in Eq.~$(\ref{eq:ActionEM})$ and
$S_j[\varphi_j] =S_j^C[\varphi_j] +S_j^T[\varphi_j]$ describes
the $j$'th junction where the Coulomb action $S_j^C$
and the tunneling action $S_j^T$ are given by Eqs.~$(\ref{eq:ActionCou})$
and $(\ref{eq:ActionTJ})$, with the
replacements $G_T \rightarrow G_j$ and $C \rightarrow C_j$. 
Performing the functional
derivatives explicitly, the current-current correlator is found to be
of the form $(\ref{eq:IICorrSJ})$ with the replacement 
$Z_{\rm JE} \rightarrow Z_{\rm AE}=Z_{\rm AE}[0]$ and the appropriate 
action $S_{\rm AE}[\{ \varphi_j \},0]$. Further, the current
functionals $I_{\rm em}[\varphi,\tau]$ now depend on the sum of phases, 
$\varphi \rightarrow \sum_{j=1}^N \varphi_j$, and the functional
integral is defined over all configurations of the phases $\varphi_j$.
As in Eq.~$(\ref{eq:G1OmegaSJ})$ the first term can be handled
exactly, and we find
$G_{\rm AE}(\omega)=Y(\omega) + G^{(2)}_{\rm AE}(\omega)$, where 
$G^{(2)}_{\rm AE}(\omega)$ given by Eq.\ $(\ref{eq:Gprime})$ and
Eq.\ $(\ref{eq:C2formal})$ with 
$F[\varphi,\nu_n] \rightarrow F[\sum_{j=1}^N \varphi_j,\nu_n]$.
So far no approximations have been made and the result follows from 
a straightforward extension of our findings for a single
junction. The
qualitative difference lies in the topological structure of the phase
configuration space and becomes clear when one evaluates the path
integral. Again, we restrict ourselves to the semiclassical limit.

\subsection{Semiclassical Limit}
Determining the classical path one has to take into account the
topological structure of the configuration space. The phases
$\varphi_j$, $j=1,\ldots ,N$ are 
canonically conjugate to the charges $Q_j$ on the junction
capacitances. Now the array has $N-1$ metallic islands in between the
junctions carrying the charges $q_j=Q_j-Q_{j+1}$ for 
$j=1, \ldots ,N-1$. These island charges are quantized in units of the
elementary charge $e$, and the phases $\psi_j$ canonically conjugate
to the $q_j$ are compact, i.e., the configuration space of the phases 
is a $(N-1)$-dimensional
torus. Accordingly, the path integral is over all configurations of
the phases $\psi_j$ with $\psi_j(\hbar \beta)=\psi_j(0)+ 2\pi k_j$
where the winding numbers $k_j$ are integers. On the other hand, 
the environmental
phase is extended and conjugate to a certain linear combination $Q$ of
the $Q_j$. Since the environment transfers charges continuously, $Q$ is
not quantized and the path integral over the environmental phase is
over all configurations with 
$\varphi_{\rm em}(0)=\varphi_{\rm em}(\hbar \beta)$. Rather
than making the canonical transform to the charges $q_j$ 
$(j=1, \ldots ,N-1)$, $Q$ and the conjugate phases explicitely, one
finds that equivalently we may integrate over all configurations of
the phases $\varphi_j$ with 
$\varphi_j(\hbar \beta)= \varphi_j(0)+ 2\pi k_j$ where the integer
winding numbers obey the constraint $\sum_{j=1}^N k_j = 0$.

At high temperatures the classical paths are
straight line flips
$\bar{\varphi}_j^{(k_j)}(\tau)=\varphi_j^0+\nu_{k_j} \tau$ running
from $\varphi_j^0$ to $\varphi_j^0+2\pi k_j$. The action is
invariant under global shifts of the $\varphi_j$ and we may set 
$\varphi_j^0=0$ for all $j=1,\ldots ,N$.  All
paths with winding number $k_j \not= 0$ are exponentially
suppressed by the classical action contribution 
$S_j^{cl} \approx \pi^2 k_j^2/\beta E_{C_j} +|k_j| G_j/2 G_K$. 
Thus, to obtain the leading order quantum corrections, we may restrict
ourselves to winding numbers $k_j=0$ for $j=1 \ldots N$. 
Finite winding numbers are considered in the next section
where we focus on the single electron transistor and go beyond the
leading order quantum correction. 

The action may
be expanded in powers of the Fourier coefficients 
$\widetilde{\varphi}_i(\nu_n)$ yielding a result of the form 
$(\ref{eq:ActionExp})$ where the second
order variational action reads
\begin{equation}
 S^0_{\rm AE}[\{\varphi_j\}]
 =
 S^0_Y \left[\sum\nolimits_{j=1}^N \varphi_j \right]
 +\sum_{j=1}^N S^0_j[\varphi_j].
\end{equation}
The environmental contribution is given by
\begin{equation}
 S^0_Y[\varphi]
 =
 \hbar \sum_{n=1}^\infty \lambda_Y(\nu_n)
 |\widetilde{\varphi}(\nu_n)|^2 ,
\end{equation}
with the eigenvalues
\begin{equation}
 \lambda_Y(\nu_n)
 =
 \frac{\hbar^2\beta}{e^2} |\nu_n| \widehat{Y}(|\nu_n|).
\label{eq:SecondAEY}
\end{equation}
The second order variational tunneling action for junction $j$
reads
\begin{equation}
 S^0_j[\varphi_j]
 =
 \hbar \sum_{n=1}^\infty
 \lambda_j(\nu_n)
 \left| \widetilde{\varphi}_j(\nu_n) \right|^2
\end{equation}
with the eigenvalues
\begin{equation}
 \lambda_j(\nu_n)
 =
 \frac{\hbar^2\beta}{e^2} |\nu_n| \widehat{G}^0_j(\nu_n),
\end{equation}
where $\widehat{G}^0_j(\nu_n )$ is given by Eq.~$(\ref{eq:GTLaplace})$ 
adapted to a junction with capacitance $C_j$ in parallel with
an Ohmic resistor $1/G_j$. The higher order variational actions are
given by straightforward extensions of Eq.~$(\ref{eq:HigherVar})$. 
Expanding in powers of the higher order terms
$S_{\rm AE}^{2k}[\varphi]$, $k=2,3,\ldots$, we are left with
expectation values of products of the phase variables 
$\widetilde{\varphi}_j(\nu_n)$.
It is now useful to define a Gaussian average
\begin{eqnarray}
 \langle X \rangle_0
 &=&
 \frac{1}{Z_{\rm AE}^0}
 \int \prod_{n=1}^\infty \prod_{j=1}^N
 d\widetilde{\varphi}_j (\nu_n ) d\widetilde{\varphi}_j^* (\nu_n )
   \nonumber              \\
 &&  
 \exp\left\{
  -\frac{1}{\hbar} S_{\rm AE}^0[\{\varphi_j\}]
 \right\}
 X
\end{eqnarray} 
with the Gaussian partition function $Z_{\rm AE}^0$ defined by the
requirement $\langle 1\rangle_0 =1$. The difference between
$Z_{\rm AE}^0$ and the full partition function $Z_{\rm AE}$ is of 
order $(\beta E_{C_j})^2$ and may be neglected here.
Due to the Gaussian form of the measure 
$\exp[- S_{\rm AE}^0[\{\varphi_j\}]/\hbar]$, the averages of products
of Fourier coefficients $\widetilde{\varphi}_j(\nu_n)$ decompose 
into sums over products of two point expectations. For two different 
phase variables $l\not= l'$ we obtain
\begin{equation}
 \langle
  \widetilde{\varphi}_l(\nu_n) \widetilde{\varphi}_{l'}(\nu_m)
 \rangle_0
 =
 - \delta_{n,-m}
 \left.
   \prod_{j \not= l,l'}^{N+1} \lambda_j(\nu_n)
 \right/  \Lambda(\nu_n)  ,
\end{equation}
with
\begin{equation}
 \Lambda(\nu_n)
 =
 \sum_{i=1}^{N+1}  \prod_{j \not= i}^{N+1} \lambda_j(\nu_n) .
\end{equation}
Here and in the remainder we define
$\lambda_{N+1}(\nu_n)=\lambda_Y(\nu_n)$ and note that summation and
multiplication indices run from $1$ if not otherwise specified. For
phase variables of the same junction we find
\begin{equation}
 \langle
  \widetilde{\varphi}_l(\nu_n) \widetilde{\varphi}_l(\nu_m)
 \rangle_0
 =
 \frac{1}{\lambda^{(l)}(\nu_n)} \delta_{n,-m},
\end{equation}
where
\begin{equation}
 \lambda^{(l)}(\nu_n)
 =
 \frac{\hbar^2 \beta}{e^2}
 |\nu_n| \widehat{G}'_l(\nu_n)
\label{eq:EigenAE}
\end{equation}
plays the role of an effective eigenvalue for phase fluctuations in
junction $l$ with all other phases $\varphi_j$, $j \not= l$ already 
traced out. Here,
\begin{equation}
 \widehat{G}'_l(\nu_n)
 =
  \widehat{G}_l^0(\nu_n) +
  \left[
   \frac{1}{\widehat{Y}(|\nu_n|)} +
   \sum_{j\not=l}^N \frac{1}{\widehat{G}_j^0(\nu_n)}
  \right]^{-1}
\label{eq:GeffAE}
\end{equation}
may be considered as the Laplace
transform of an effective response function 
describing the circuit seen from junction $l$, {\it i.e.}, a series of
$N-1$ junctions and an environmental impedance in parallel to 
junction $l$ where the junctions are described effectively 
by linear elements. Including the
fourth order variational derivative of the action, we get as a
generalization of Eq.~$(\ref{eq:C_2})$
\begin{eqnarray}
 F_{\rm AE}(\nu_n)
 &=&
 \frac{4e^2\beta}{\hbar}\,
 \frac{\widetilde k(\nu_n)^2}{\Lambda(\nu_n)}
 \sum_{l=1}^N \prod_{j\not=l}^N \lambda_j(\nu_n)
 \Bigg[
   1         \nonumber \\ 
 &&
 + \frac{G_l}{G_K}
   \frac{2\beta}{\Lambda(\nu_n)} \prod_{j\not=l}^N \lambda_j(\nu_n)
          \nonumber \\ 
 && \times
   \sum_{{m=-\infty}\atop {m\ne 0}}^{\infty}
   \frac{\widetilde{\alpha}(\nu_{n+m})
          -\widetilde{\alpha}(\nu_n)
          -\widetilde{\alpha}(\nu_m)}{\lambda^{(l)}(\nu_m)}
 \Bigg]  .
 \label{eq:C2explA}
\end{eqnarray}
Now, the convergence of the expansion depends on the effective
eigenvalues $(\ref{eq:EigenAE})$. To estimate the range of validity, 
we consider the smallest eigenvalues which at high temperatures are
given by
$\lambda^{(l)}(\nu_1) \approx 2 \pi^2 / \beta E_{C_l}$. 
Again the analytic continuation gives rise 
to a reduction of the order of the quantum corrections in the
expansion parameter. For contributions with vanishing winding number 
the arguments given in the previous
section apply likewise to the present problem.
On the other hand,
for lower temperatures one has to take
into account winding numbers $k_j \not= 0$ and finds that some of
the eigenvalues tend to zero. The marginally stable fluctuation modes
lead to a breakdown of the simple semiclassical 
treatment. The appropriate extension of the semiclassical
approximation was discussed elsewhere \cite{WangBOXPRB96}. 
The topological structure of the phase space leading
here to a breakdown of the simple semiclassical approximation  
at low temperatures even for large conductance is the main difference
between the single tunnel junction with environment and circuits
containing many junctions.
As a result one finds that the truncated expression 
$(\ref{eq:C2explA})$
is valid up to first order in
$\epsilon=\mbox{Max}\left( \beta E_{C_j}: j=1,\ldots ,N \right)$.
After the analytical continuation 
$\nu_n \rightarrow -i\omega + \delta$ we can write the total 
conductance in the compact form
\begin{equation}
 G_{\rm AE}(\omega)
 =
 \left[
  \frac{1}{Y(\omega)}+\sum_{j=1}^N \frac{1}{G_{\rm eff}^j(\omega)}
 \right]^{-1}  ,
\label{eq:condAE}
\end{equation}
describing $N+1$ linear elements in series: $G_{\rm eff}^j(\omega)$
with $j=1, \ldots ,N$ and the admittance $Y(\omega)$.
Here, the $G_{\rm eff}^j(\omega)$ are of the form 
$(\ref{eq:EffLinSJ})$ where the
auxiliary functions ${\cal U}_j$ are given by 
Eq.~$(\ref{eq:qmsuppSJ})$ with
$\lambda_{\rm JE}$ replaced by $\lambda^{(j)}$ introduced in 
Eq.~$(\ref{eq:EigenAE})$. 
This is a straightforward extension of the result in the previous 
section valid to
linear order in $\epsilon$ for arbitrary conductances $G_j$ and
admittances $Y(\omega)$.

\subsection{Discussion of Results}
For a more explicit discussion of the results we consider now $N$
identical junctions $G_j=G$ and $C_j=C$. The
eigenvalues $(\ref{eq:EigenAE})$ then read
%
\begin{eqnarray}
 &&\lambda(\nu_n) 
 =
 \lambda^{(j)}(\nu_n)
     \nonumber             \\
 &&
 =
 \frac{\hbar^2 \beta}{e^2}
 |\nu_n|
 \left[
  \widehat{G}_j^0(\nu_n) +
   \frac{ \widehat{Y}(|\nu_n|) \widehat{G}_j^0(\nu_n) }
        { (N-1) \widehat{Y}(|\nu_n|) + \widehat{G}_j^0(\nu_n) }
 \right]  
\label{eq:lambdaNequal}
\end{eqnarray}
%
and coincide for all junctions. For the total conductance
of the array $(\ref{eq:condAE})$ we obtain
\begin{equation}
 G(\omega)
 =
 \frac{G_{\rm eff}(\omega) Y(\omega)}
      {G_{\rm eff}(\omega) + N Y(\omega)}.
\label{eq:GallgAJ}
\end{equation}
where 
\begin{equation}
 G_{\rm eff}(\omega)
 =
 G[1-{\cal U}(\omega)]-i \omega C
\end{equation}
is the effective linear conductance of one junction.
In this order each junction can be described by a linear element
$G^*(\omega)=G[1-{\cal U}(\omega)]$, depending on the
circuit, in parallel with the geometrical capacitance $C$ as depicted
in Fig.~\ref{fig:fig2}a. To proceed we consider an Ohmic
environment $Y(\omega)=Y$ and find for the effective linear
element
\end{multicols} \widetext
\begin{eqnarray}
 \frac{G^*(\omega)}{G}
 &=&
 1-
 \left\{
  \frac{(N-1)[\psi(1+u_T+\tilde{\omega})
        -\psi(1+\tilde{\omega})] }{u_T}
  +
  \frac{\psi(1+u_N+\tilde{\omega})
        -\psi(1+\tilde{\omega}) }{u_N}
 \right.
 \nonumber    \\
&& 
  \left.
  +
  \frac{(N-1)[\psi(1+u_T+\tilde{\omega})
        -\psi(1+u_T)] +
         \psi(1+u_N+\tilde{\omega})
        -\psi(1+u_N)}
       {\tilde{\omega}}
 \right\} \frac{\beta E_C}{\pi^2 N},
\label{eq:ohmomegaA}
\end{eqnarray}
\begin{multicols}{2} \narrowtext
where 
\begin{equation}
 u_N
 =
 \frac{G+NY}{G_K} \frac{\beta E_C}{2\pi^2}
 , \quad
 u_T
 =
 \frac{G}{G_K} \frac{\beta E_C}{2\pi^2}
 , \quad
 \tilde{\omega}=\frac{\hbar \beta}{2 \pi i} \omega
\end{equation}
are auxiliary quantities and $E_C=e^2/2C$ is the charging energy for
{\it one} junction. For $N=1$ we recover the results of Sec.~III, 
of course. For a large array with $N \gg 1$, terms in
Eq.~$(\ref{eq:ohmomegaA})$ containing $u_N$ drop out, and the quantum
suppression becomes independent of $Y$. 
Furthermore the high-temperature anomaly, 
{\it cf}.\ Fig.~\ref{fig:fig4}, is now a $1/N$ effect and the limiting
result for $N \rightarrow \infty$ is analytic.

\begin{figure}
\begin{center}
\leavevmode
\epsfxsize=0.45 \textwidth
\epsfbox{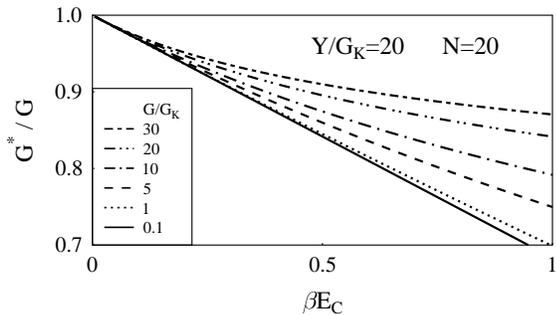}
\end{center}
\caption{Renormalized conductance $G^*/G$ of an array of
$N=20$ tunnel junctions in dependence of $\beta E_C$ for $Y/G_K=20$
and various tunneling conductances $G/G_K$.}
\label{fig:fig9}
\end{figure}

For small frequencies the
effective element behaves like an Ohmic resistor $1/G^*(\omega=0)$ with
a renormalized capacitance in parallel. The dc conductance is given by
\begin{eqnarray}
 && \frac{G^*(\omega=0)}{G}
     \nonumber             \\
 &&
 =
 1-
 \left\{
  (N-1)\left[
  \frac{\gamma+\psi(1+u_T)}{u_T} + \psi'(1+u_T)
       \right]  
 \right. 
   \nonumber   \\
&& \quad \qquad
 \left.
  +\frac{\gamma+\psi(1+u_N)}{u_N} + \psi'(1+u_N)
 \right\} \frac{\beta E_C}{N \pi^2}.
\label{eq:ArrayOhm0}
\end{eqnarray}
For $N=2$ and 
$Y/G_K \rightarrow \infty$ this reduces to
\begin{equation}
 \frac{G^*}{G}
 =
 1-
  \left[
   \frac{\gamma+\psi(1+u_T)}{u_T} + \psi'(1+u_T)
  \right]
  \frac{\beta E_C/2}{\pi^2}
\label{eq:ArraySET}
\end{equation}
coinciding with earlier findings for the symmetrical
SET \cite{ZaikinSETJETP96,GeorgSETPRB98}.
On the other hand, in the limit of large $N$ and moderate 
$G {\tiny{\lower 2pt \hbox{$<$} \atop \raise 5pt \hbox{$\sim$}}} G_K$, 
Eq.~$(\ref{eq:ArrayOhm0})$ reduces to
\begin{equation}
 \frac{G^*}{G}
 =
 1- \frac{N-1}{N} \frac{\beta E_C}{3}
\label{eq:PekolaLimitAE}
\end{equation}
again in accordance with earlier findings 
\cite{PekolaArrayJLTP97} derived from rate
theory for small tunneling conductances.
To discuss the strong tunneling corrections, 
we show in Fig.~\ref{fig:fig9} the renormalized conductance of 
an array of
$N=20$ tunnel junctions in dependence on $\beta E_C$ for $Y/G_K=20$
and various tunneling conductances $G/G_K$.
We find that the weakly conducting case $G/G_K=0.1$ perfectly coincides
with the limiting formula $(\ref{eq:PekolaLimitAE})$ (both depicted by
the solid line) whereas for larger tunneling conductances 
strong deviations from this behavior appear.

\begin{figure}
\begin{center}
\leavevmode
\epsfxsize=0.45 \textwidth
\epsfbox{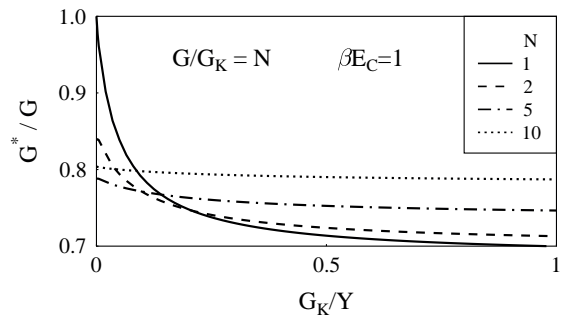}
\end{center}
\caption{Renormalized conductance $G^*/G$ of an array of
$N=1,2,5$, and $10$ equivalent tunnel junctions leading to the same
classical series conductance for $\beta E_C=1$ as a function
of the inverse environmental conductance $G_K/Y$.}
\label{fig:fig10}
\end{figure}

Experimentally one is interested in the dependence on the array
length at fixed classical series conductance. In 
Fig.~\ref{fig:fig10} we show $G^*/G$ for various $N$ whereby $G$
increases with the array length to keep the total classical series
conductance constant. Whereas for $N<5$ the renormalized 
conductance depends strongly on $Y$, it becomes
independent for large $N$.

\begin{figure}
\begin{center}
\leavevmode
\epsfxsize=0.4 \textwidth
\epsfbox{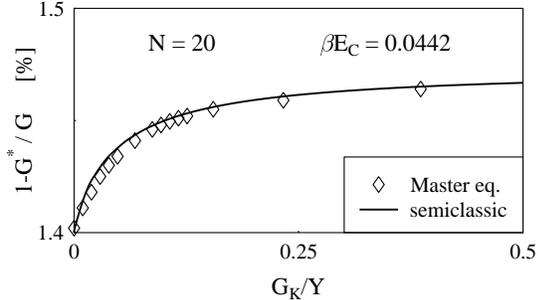}
\end{center}
\caption{Zero bias dip $1-G^*/G_T$ in per cent for an array of 
length $N=20$
and $\beta E_C=0.0442$ as a function of the inverse environmental 
conductance $G_K/Y$ in the perturbative limit compared with a
numerical master equation approach by Farhangfar 
{\it et al}.~\protect\cite{Farhangfar99}. }
\label{fig:fig11}
\end{figure}

The comparison with available experimental data 
\cite{Farhangfar99} is complicated by the large number of 
parameters, in particular, the charging energy differs from sample 
to sample with different array length. To test our predictions at
least in the perturbative regime, we compare  
with the results of a master equation 
approach \cite{Farhangfar99} based on the 
$P(E)$ theory \cite{IngoldNato92}. In Fig.~\ref{fig:fig11} we show the
zero bias dip $1-G^*/G_T$ in per cent for an array of length $N=20$ and 
$\beta E_C=0.0442$ in the limit $G \rightarrow 0$. We find good
agreement between the numerical calculations by Farhangfar 
{\it et al}.~\cite{Farhangfar99} 
and the analytical semiclassical result $(\ref{eq:ArrayOhm0})$.

The renormalized capacitance $C_{\rm eff}$ includes the linear part in
$\omega$ of
$G^*(\omega)$ and the geometrical capacitance $C$ and reads
\begin{eqnarray}
 \frac{C_{\rm eff}}{C}
 \! &=&
 1+ \frac{G}{G_K}
  \nonumber \\
 && \times
 \left\{
  (N-1)\left[
  \frac{\frac{\pi^2}{3} -2\psi'(1+u_T)}{u_T}
  -\psi''(1+u_T)
     \right] 
 \right.
  \nonumber \\
 && \!\!
 \left.
  +\frac{\frac{\pi^2}{3} -2\psi'(1+u_N)}{u_N}
  \! -\psi''(1+u_N) \!
 \right\} \!
 \frac{(\beta E_C)^2}{4 N \pi^4}
\label{eq:ArrayCap0}
\end{eqnarray}
showing a quadratic dependence on $\beta E_C$. The renormalization
is suppressed at high temperatures and also vanishes linearly for
large conductance in accordance with the behavior of a 
single tunnel junction with environment.

\section{Single Electron Transistor}

\subsection{Generating Functional and Conductance}
\begin{figure}
\begin{center}
\leavevmode
\epsfxsize=0.4 \textwidth
\epsfbox{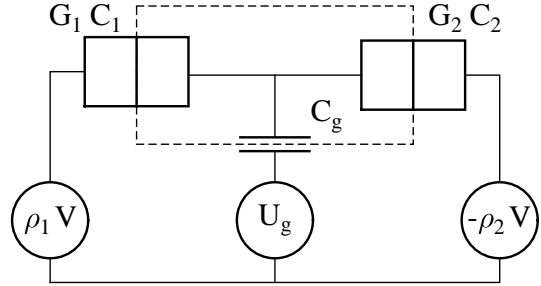}
\end{center}
\caption{Circuit diagram of the single electron transistor.}
\label{fig:fig12}
\end{figure}

The SET consists of two tunnel junctions with
tunneling conductances $G_1$, $G_2$ and capacitances $C_1$, $C_2$,
respectively, biased by a
voltage source $V$, cf.\ Fig.~\ref{fig:fig12}. The voltage may be split
among the branches in $\rho_1 V$ and $\rho_2 V$ with $(\rho_1+\rho_2=1)$. 
The island in
between the junctions is connected via a gate capacitance $C_g$ to a
control voltage $U_g$ shifting the electrostatic energy of the
system continuously. The important energy scale is the charging
energy $E_C=e^2/2C$ with the island capacitance $C= C_1+ C_2 +C_g$. For
weak electron tunneling, $E_C$ is the energy needed to charge the
island with one excess electron at vanishing gate voltage $U_g=0$.
Due to the periodicity of the Hamiltonian in $U_g$, the
conductance is a periodic function with period 1 of the
dimensionless gate voltage $n_g=U_g C_g/e$ \cite{Averin91}.

\noindent 
Following the lines of reasoning 
in the previous sections, we start with the
Lagrangian description with phases $\varphi_1, \varphi_2$ across the 
tunnel junctions and $\varphi_g$ across the gate capacitor. 
Here, we treat the shunt capacitor $C_g$ explicitely and do not 
introduce an effective environmental impedance. For the
circuit depicted in Fig.~\ref{fig:fig12} there are two independent
circuit loops leading to the constraints 
$\varphi_1 - \varphi_2 - \psi = const.$ and 
$\varphi_1 - \varphi_g - \psi_g - \rho_1 \psi = const.$, where $\psi$ is
the phase $(\ref{eq:PhasePsi})$ of the transport voltage $V$ and $\psi_g$ 
the corresponding phase of the gate voltage $U_g$. Eliminating
$\varphi_2$ and $\varphi_g$ in favor of $\varphi \equiv \varphi_1$,
we find
\begin{eqnarray}
 H'_{\rm SET}
 &=&
 H_C(
   Q + e n_g + V(C_2+\rho_1 C_g)
    ) 
     \nonumber             \\
 &&
 +  H^{(1)}_{\rm T}(\varphi )  
 +  H^{(2)}_{\rm T}(\varphi - \psi)
 +  H^{(1)}_{\rm qp} + H^{(2)}_{\rm qp} .
\label{eq:HamSETwo}
\end{eqnarray}
The Coulomb Hamiltonian $H_C$ is given by Eq.~$(\ref{eq:HamCou})$ 
and the other terms are defined by 
Eqs.~$(\ref{eq:HamQP})$ and $(\ref{eq:HamTun})$ 
with corresponding labels.
After a unitary transformation one finds equivalently 
\begin{eqnarray}
 H_{\rm SET}
 &=&
 H_C(Q + e \widetilde{n}_g)  
 +  H^{(1)}_{\rm T}(\varphi + \kappa_1 \psi)
     \nonumber             \\
 &&  
 +  H^{(2)}_{\rm T}(\varphi - \kappa_2 \psi)
 +  H^{(1)}_{\rm qp} + H^{(2)}_{\rm qp}
\label{eq:HamSET}
\end{eqnarray}
with an arbitrary shift parameter $\kappa_1$ and
$\kappa_2=1-\kappa_1$. Here we introduced a shifted dimensionless 
gate voltage 
\begin{equation}
 e \widetilde{n}_g
 =
 U_g C_g + V (C_2 + \rho_1 C_g - \kappa_1 C) .
\end{equation}
In the sequel we restrict ourselves to zero 
frequency where the expectation
values of the current operators
through both junctions coincide 
$\langle I_1 \rangle(\omega=0)=\langle I_2 \rangle(\omega=0)$. 
The first current operator
$I^{(1)}$ may then be chosen as an arbitrary linear combination
$I^{(1)}=\epsilon_1 I_1-\epsilon_2 I_2$ (with
$\epsilon_1+\epsilon_2=1$) of the tunneling current operators
$I_1$ and $I_2$ through junctions $1$ and $2$, respectively. The
relative minus sign comes from the opposite directions of $I_1$
and $I_2$, which are both positive for flux onto the island. The
second current operator $I^{(2)}=\kappa_1 I_1 -\kappa_2 I_2$ 
is determined as above by the linear coupling term
to the transport voltage $V$.
The dependence of the Coulomb Hamiltonian on the transport 
voltage may be removed by a gate voltage shift and thus need not
be considered. 
Moreover, this coupling would lead to a
displacement current contribution vanishing at $\omega =0$.
To evaluate the
current-current correlator we employ the generating functional
$(\ref{eq:genfunc})$. The current operators through the junctions 
are given by Eq.~$(\ref{eq:TunCurrent})$, for $j=1,2$.
To derive the generating functional we define
\begin{equation}
 \widetilde{H}^{(j)}_{\rm T} 
 =
 H^{(j)}_{\rm T}(\varphi) 
 - I_j \xi_j(\tau)  .
\label{eq:HamTildeT}
\end{equation}
The new tunneling Hamiltonians are of the form $(\ref{eq:HamTun})$ 
where $\Lambda$ is replaced
by $[1 + ie\xi_j(\tau)/\hbar]\Lambda_j$. 
With these replacements we get for
the generating functional
\begin{equation}
 Z_{\rm SET}[\xi_1,\xi_2]
 =
 \int D[\varphi] \exp
  \left\{
   -\frac{1}{\hbar} S_{\rm SET}[\varphi,\xi_1,\xi_2]
  \right\},
\end{equation}
where the effective action reads
\begin{equation}
 S_{\rm SET}[\varphi,\xi]
 =
 S_{\rm SET}^C[\varphi]
 +S_1^T[\varphi,\xi_1]
 +S_2^T[\varphi,\xi_2].
\end{equation}
The first term on the rhs
\begin{equation}
 S_{\rm SET}^C[\varphi]
 =
 \int_0^{\hbar \beta}d\tau
 \left[
  \frac{\hbar^2 {\dot \varphi}^2(\tau)}{4 E_C}
  + i \hbar n_{\rm g}{\dot \varphi}(\tau)
 \right]
\end{equation}
describes Coulomb charging of the island in presence of an
applied gate voltage. The effective tunneling action
\begin{eqnarray}
 S_j^T[\varphi,\xi_j]
 \! &=& \!
 -\frac{G_j}{G_K} \!
 \int_0^{\hbar \beta} \! d\tau 
 \int_0^{\hbar \beta} \! d\tau'
 \alpha(\tau-\tau') 
 \left[
   1-i\frac{e}{\hbar}\xi_j(\tau)
 \right]
     \nonumber             \\
 && \times
 \left[
   1+i\frac{e}{\hbar}\xi_j(\tau')
 \right]
 e^{i[\varphi(\tau)-\varphi(\tau')]}
\end{eqnarray}
describes quasi-particle tunneling through junction $j$ 
with the kernel
$\alpha(\tau)$ given by Eq.\ $(\ref{eq:AlphaFou})$. For vanishing
auxiliary field $\xi_j = 0$, the action reduces to the 
single electron box action \cite{WangBOXPRB96} 
$S_{\rm SET}[\varphi]= S_{\rm SET}[\varphi,0,0]=S_{\rm box}$ 
depending only on the parallel
conductance $G_{||} = G_1+G_2$. One has
\begin{eqnarray}
 S_1^T[\varphi,0] + S_2^T[\varphi,0]
 &=&
 2 \frac{G_{||}}{G_K}
 \int_0^{\hbar \beta} d\tau \int_0^{\hbar \beta} d\tau'
     \nonumber             \\
 &&    \!\!
 \alpha(\tau-\tau')
 \sin^2
 \left[
  \frac{\varphi(\tau)-\varphi(\tau')}{2}
 \right]  .
\label{eq:SBox}
\end{eqnarray}
Thus, the generating functional at vanishing auxiliary fields gives
the box partition function 
$Z_{\rm SET}=Z_{\rm SET}[0,0]=Z_{\rm box}$. 
Performing the variational derivatives explicitly, we get
for the correlator
\begin{equation}
 \langle
  I_j(\tau) I_{j'}(\tau')
 \rangle
 =
 \langle 
  I_j(\tau) I_j(\tau')
 \rangle^E \delta_{j,j'} +
 \langle 
  I_j(\tau) I_{j'}(\tau')
 \rangle^F.
\label{eq:corr0}
\end{equation}
Since the auxiliary fields are in the argument of an exponential, there
are two contributions. The first term comes from the second order
variational derivative of the action and reads 
\begin{eqnarray}
 \langle 
  I_j(\tau) I_j(\tau')
 \rangle^E
 &&=
 4\pi G_j \alpha(\tau-\tau') 
 \frac{1}{Z_{\rm SET}}
 \int D[\varphi]  
     \nonumber             \\
 &&    
 \exp 
  \! \left\{\! 
    -\frac{1}{\hbar}S_{\rm SET}[\varphi] 
  \!\right\}
  \cos[\varphi(\tau)-\varphi(\tau')] .
\label{eq:corr1}
\end{eqnarray} 
The second term in Eq.\ $(\ref{eq:corr0})$ involves a 
multiplication of two current
functionals arising as first order variational derivatives of the action
\begin{eqnarray}
 \langle 
  I_j(\tau) I_{j'}(\tau')
 \rangle^F
 &&=
 \frac{G_j G_{j'}}{G_K^2} \frac{1}{Z_{\rm SET}}
 \int D[\varphi]  
     \nonumber             \\
 &&  
  \exp\left\{-\frac{1}{\hbar}S_{\rm SET}[\varphi]\right\}
  I[\varphi,\tau] I[\varphi,\tau'],
\end{eqnarray}
with the current functional
\begin{equation}
 I[\varphi,\tau]
 =
 \frac{2e}{\hbar} 
 \int_0^{\hbar\beta}d\tau'
 \alpha(\tau-\tau')\sin[\varphi(\tau)-\varphi(\tau')].
\end{equation}
Taking into account that 
$\langle I_j(\tau) I_{j}(\tau') \rangle^E/G_j$ and 
$\langle  I_j(\tau) I_{j'}(\tau')  \rangle^F/ G_j G_{j'}$ 
depend only on the parallel
conductance $G_{||}=G_1+G_2$ and thus are
independent of the indices $j$ and $j'$, the conductance may be
written as 
\begin{eqnarray}
 G
 &=&
    \epsilon_1\kappa_1 ( G_1 E + G_1^2 F) 
  - (\epsilon_1 \kappa_2 + \epsilon_2 \kappa_1 )  G_1 G_2 F
     \nonumber             \\
 &&
  + \epsilon_2\kappa_2 ( G_2 E + G_2^2 F), 
\end{eqnarray}
where
\begin{equation}
 E
 =
 \lim_{\omega \rightarrow 0} 
 \frac{1}{\hbar \omega} {\rm Im}
    \lim_{i \nu_l \rightarrow \omega + i\delta}
    \int_0^{\hbar \beta} \! d\tau \, e^{i\nu_l \tau}
    \frac{   
      \langle I_1(\tau) I_1(0) \rangle^E 
         }{G_1}
\end{equation}
and
\begin{equation}
 F
 =
 \lim_{\omega \rightarrow 0} 
 \frac{1}{\hbar \omega} {\rm Im}
    \lim_{i \nu_l \rightarrow \omega + i\delta}
    \int_0^{\hbar \beta} \! d\tau \, e^{i\nu_l \tau}
    \frac{
       \langle I_1(\tau) I_1(0) \rangle^F 
         }{G_1^2} .
\end{equation}
Since the conductance does not depend on
the specific choice of the parameters $\epsilon_j$ and $\kappa_j$, we
then find that
\begin{equation}
G_{\rm SET}=G_{\rm cl} E,
\label{eq:kubo2}
\end{equation}
with the classical series conductance
\begin{equation}
G_{\rm cl}=\frac{G_1 G_2}{G_1+G_2}.
\label{eq:Gclass}
\end{equation}
This is a formally exact expression for the linear dc conductance. 
To proceed, we make explicit the sum over winding numbers $k$ of the
phase and write the correlator (\ref{eq:corr1}) in the form
\end{multicols} \widetext
\begin{equation}
 \langle
  I_1(\tau) I_1(0)
 \rangle^E /G_1
 =
  4\pi \alpha(\tau) \frac{1}{Z_{\rm SET}}
  \sum_{k=-\infty}^{\infty}
  \int\limits_{\varphi(0)=0}^{\varphi(\hbar\beta)=2\pi k}
  D[\varphi]      
  \exp\left\{ -\frac{1}{\hbar}S_{\rm SET}[\varphi] \right\} 
  \cos[\varphi(\tau)-\varphi(0)]. 
\label{eq:korrSET}
\end{equation}
\begin{multicols}{2} \narrowtext
This result may
be used as a starting point for analytical work and/or
numerical calculations \cite{GeorgMCSVD99}. For further
analysis, here we consider the semiclassical approximation.

\vspace{1cm}

\subsection{Semiclassical limit}
For given winding number $k$, the path integral may be evaluated
approximately by expanding around the classical paths
$\bar{\varphi}^{(k)}(\tau)=\varphi^0+\nu_k \tau$. An arbitrary
path of winding number $k$ may be written
$\varphi(\tau)=\bar{\varphi}^{(k)}(\tau)+\zeta(\tau)$ with
$\zeta(0)=\zeta(\hbar \beta)=0$. In terms of the Fourier
coefficients $\widetilde{\zeta}(\nu_n)$ the action reads
\begin{equation}
 S_{\rm SET}[\bar{\varphi}^{(k)}+\zeta]
 =
 2\pi i k \hbar n_g +
 S_{\rm SET}^{(k)}[\zeta]
 + \sum_{m=2}^\infty \delta^m S_{\rm SET}^{(k)}[\zeta],
\label{eq:ActExpSET}
\end{equation}
where the first term on the rhs is the topological
contribution and
\begin{equation}
 S_{\rm SET}^{(k)}
 =
 \hbar \left(
 \frac{\pi^2 k^2}{\beta E_C}+|k|\frac{g}{2}
 \right)
\end{equation}
the classical action of winding number $k$,
with the dimensionless parallel conductance $g=G_{||}/G_K$. 
The second order variational action
\begin{equation}
 \delta^2S_{\rm SET}^{(k)}
 =
 \hbar \sum_{n=1}^\infty \lambda_{\rm SET}^{(k)}(\nu_n)
 \left|
   \widetilde{\zeta}(\nu_n)
 \right|^2
\end{equation}
is diagonal with the eigenvalues
\begin{equation}
 \lambda_{\rm SET}^{(k)}(\nu_n)
 =
 \frac{2\pi^2n^2}{\beta E_C}+g \Theta(n-|k|)(n-|k|).
\end{equation}
The higher order terms in $(\ref{eq:ActExpSET})$ read
\begin{eqnarray}
 \delta^{(2m+1)} S_{\rm SET}^{(k)}
 &=&
 g \frac{(-1)^m}{(2m+1)!}
 \int_0^{\hbar \beta} d\sigma d\sigma'
 \alpha(\sigma-\sigma') 
     \nonumber             \\
 && 
 \sin[\nu_k(\sigma-\sigma')]
 [\zeta(\sigma)-\zeta(\sigma')]^{2m+1}
\end{eqnarray}
for odd orders and
\begin{eqnarray}
 \delta^{(2m)} S_{\rm SET}^{(k)}
 &=&
 g \frac{(-1)^{m+1}}{(2m)!}
 \int_0^{\hbar \beta} d\sigma d\sigma'
 \alpha(\sigma-\sigma') 
     \nonumber             \\
 && 
 \cos[\nu_k(\sigma-\sigma')]
 [\zeta(\sigma)-\zeta(\sigma')]^{2m}
\end{eqnarray}
for even orders, with $m=1,2,\ldots$. 
Since $\lambda_{\rm SET}^{(k)}(\nu_n)$ is
large for small $\beta E_C$, the expansion $(\ref{eq:ActExpSET})$
about the classical path converges rapidly for high temperatures.
At low temperatures $\lambda_{\rm SET}^{(k)}(\nu_n)$ 
vanishes for $n<|k|$ and the
simple semiclassical approximation breaks down. 
The zero modes can be treated systematically
for large $g$ by considering quasi-classical
trajectories with collective coordinates (sluggons) 
and fluctuations around them. This treatment, presented elsewhere
\cite{WangBOXPRB96} for the partition function of
the single electron box, 
lies outside the scope of the present work and we proceed with
the high temperature expansion. Rewriting the cosine function in Eq.\
$(\ref{eq:korrSET})$ as a sum of exponentials, we may perform the path
integral and get for the correlator
\begin{eqnarray}
 &&  \!\!\!\!\!\!\!
 \langle
  I_1(\tau) I_1(0)
 \rangle^E /G_1 
     \nonumber             \\
 &=& 
 4\pi \alpha(\tau) \frac{1}{Z_{\rm SET}}
 \sum_{k=-\infty}^{\infty} C_{|k|} e^{-2\pi i k n_g}
    \nonumber \\
&&
 \exp
 \left[-2\sum_{m=1}^\infty
  \frac{1-\cos(\nu_m \tau)}{\lambda_{\rm SET}^{(k)}(\nu_m)}
 \right]
 \left[1- \frac{1}{\hbar} S_4^{(k)}+... \right],
\label{eq:korr2}
\end{eqnarray}
where the coefficients $C_k$ read \cite{WangBOXPRB96}
\begin{equation}
C_k=\frac{\Gamma(1+k_+)\Gamma(1+k_-)}{\Gamma^2(1+k)\Gamma(1+u)}
 e^{-S_{\rm SET}^{(k)}},
\end{equation}
with $k_\pm=k+\frac{u}{2}\pm\frac{1}{2}\sqrt{4uk+u^2}$ and $u=g
\beta E_C/2\pi^2$. The contribution of the third order variational 
action cancels, thus the dominant correction to the semiclassical
approximation stems from the fourth order term
\end{multicols} \widetext
\begin{equation}
 S_4^{(k)}
 =
 \frac{1}{2} g \hbar \beta
  \sum_{{m,l=-\infty} \atop {m,l\not=0}}^\infty
  \frac{
    {\widetilde \alpha}(\nu_k)
    -2{\widetilde \alpha}(\nu_{l+k})-2{\widetilde \alpha}(\nu_{l-k})
    +{\widetilde \alpha}(\nu_{m+l+k})+{\widetilde \alpha}(\nu_{m+l-k})
  }{\lambda_{\rm SET}^{(k)}(\nu_m) \lambda_{\rm SET}^{(k)}(\nu_l)}.
\label{eq:fourth}
\end{equation}
\begin{multicols}{2} \narrowtext
The corresponding expansion of the partition function 
$Z_{\rm SET}$ reads
\begin{equation}
 Z_{\rm SET}
 =
 \sum_{k=-\infty}^\infty C_{|k|} e^{2\pi ik n_g}
 \left[1- \frac{1}{\hbar} S_4^{(k)}+... \right],
\label{eq:partition}
\end{equation}
with the same correction
$(\ref{eq:fourth})$. The expansions
$(\ref{eq:korr2})$ and $(\ref{eq:partition})$ proceed in powers of
$\beta E_C$, however, terms involving $u=g\beta E_C/2\pi^2$ are kept
to {\it all} orders. This ensures a meaningful result in the limit
of moderately high temperatures also for large parallel conductance $g$.

From Eq. $(\ref{eq:korr2})$ one obtains for the Fourier coefficients
\begin{equation}
 E(\nu_n)
 =
 \int_0^{\hbar \beta} d\tau \, e^{i\nu_n \tau}
    \langle I_1(\tau) I_1(0) \rangle^E /G_1
\end{equation}
the high temperature expansion
\begin{eqnarray} 
 E(\nu_n) \!
 &= \!&
 \frac{4 \pi}{Z_{\rm SET}}
 \sum_{k=-\infty}^\infty
  C_{|k|} e^{-2\pi i k n_g}
     \nonumber             \\
 &&
  \exp
  \left[ 
    -2\sum_{l=1}^\infty \frac{1}{\lambda_{\rm SET}^{(k)}(\nu_l)} 
  \right]
  \Bigg\{
   {\widetilde \alpha}(\nu_{n+k})
            \nonumber  \\
&& 
   + 
   \sum_{m \not=0}
    \frac{{\widetilde \alpha}(\nu_{n+k+m})
         }{\lambda_{\rm SET}^{(k)}(\nu_m)}+
    \frac{1}{2}\sum_{m,l \not=0}
    \frac{{\widetilde \alpha}(\nu_{n+k+m+l})
         }{\lambda_{\rm SET}^{(k)}(\nu_m)
           \lambda_{\rm SET}^{(k)}(\nu_l)
          } 
     \nonumber             \\
 &&
    -\frac{1}{\hbar}
     {\widetilde \alpha}(\nu_{n+k})S_4^{(k)}
    + {\cal O}(\beta E_C)^3
 \Bigg\}.
\label{eq:fourier}
\end{eqnarray}
Since no $1/\lambda_{\rm SET}^{(k)}(\nu_n)$ term appears,
the order of the expression remains the same after 
analytical continuation.
When $E(\nu_n)$ is analytically continued in the complex
$\nu$ plane, $E(\nu)$ is analytic on each half plane
Re$\,\nu{\tiny{\lower 2pt \hbox{$<$} \atop \raise 5pt
\hbox{$>$}}}0$ with a cut along the imaginary axis \cite{BaymGreensJMP61}.
The representation of $E(\nu)$ as a sum over winding
numbers $k$ shifts this cut to Re$\,\nu=k$ for the $k$'th term of
the sum. Thus, in the phase representation, only the full sum
shows the analytic properties underlying the conductance formula
$(\ref{eq:Kubo})$. The summands of the high temperature expansion
$(\ref{eq:fourier})$ for winding number $k$ are of the form
$g(k)f(|n+k|)=\Theta(-n-k)g(k)f(-n-k)+\Theta(n+k)g(k)f(n+k)$, 
where $f(n),g(n)$ are analytic functions and $g(k)=g(-k)$. 
Now, the sum over winding numbers may be
expressed as a sum over charges $m$
\begin{eqnarray}
 \sum\limits_{k=-\infty}^\infty & g(k) & f(|n+k|)
     \nonumber             \\
 &=&
 \sum_{m=-\infty}^{\infty}
 \int_{-\infty}^{-n} d\kappa\, e^{2\pi im\kappa} g(\kappa)f(-n-\kappa)
     \nonumber             \\
 &&
 + 
 \int_{-n}^\infty d\kappa \, e^{2\pi im\kappa} g(\kappa)f(n+\kappa).
\end{eqnarray}
Performing the limits 
$i\nu_n \rightarrow \omega +i \delta$ and $\omega \rightarrow 0$
for the
rhs of this equation and rewriting the result again as a sum over
winding numbers, we obtain
\begin{eqnarray}
&&
 \lim\limits_{\omega \rightarrow 0}\frac{1}{\hbar \omega}
  {\rm Im}  \lim\limits_{i\nu_n \rightarrow \omega +i \delta} 
 \sum_{k=-\infty}^\infty g(k)f(|n+k|)
     \nonumber             \\
 &&
 =
 -\frac{\hbar\beta}{2\pi}\sum_{k=-\infty}^\infty
  g(k) \frac{\partial}{\partial k} f(|k+\delta|)
 =
 -\frac{\hbar\beta}{2\pi}\, g(0)f'(0).
\end{eqnarray}
The sums in Eq.~$(\ref{eq:fourier})$ may be performed exactly 
with the help of integral representations of the psi function 
and its derivatives
\cite{Abramowitz70}
\begin{equation}
 \psi(z)
 =
 \int_0^\infty dt
 \left(
  \frac{e^{-t}}{t} - \frac{e^{-zt}}{1-e^{-t}}
 \right)
\end{equation}
and
\begin{equation}
 \psi^{(n)}(z)
 =
 (-1)^{n+1} \int_0^\infty dt
  \frac{t^n e^{-zt}}{1-e^{-t}} .
\end{equation}
This way the high temperature expansion of the conductance
may be evaluated to read
\begin{eqnarray}
 G_{\rm SET}
 \! &=& \!
  G_{\rm cl} Z_{\rm SET}^{-1}
  \exp\left\{-2[\gamma+\psi(1+u)]/g \right\}
  \nonumber \\
 && \!
  \big\{
    1 - \psi'(1+u) (\beta E_C/ \pi^2)
  \nonumber \\
 &&
    \!+ \! \left[ g \sigma(u)  + \tau(u)  \right]
      (\beta E_C/ 2 \pi^2)^2
    \!+{\cal O}(\beta E_C)^3
  \big\}   .
\label{conductance}
\end{eqnarray}
The dependence on $u=g\beta E_C/2\pi^2$ is given in terms of
two auxiliary functions
\begin{eqnarray}
 \sigma(u)
 &=&
  \frac{\gamma+\psi(1+u)-u\psi'(1+u)}{u^2}
     \nonumber             \\
 &&
 + \int_0^1 \! \mbox{d}v \, \frac{2v(1-v^u)\phi(v,1,1+u)}{(1-v)u}
\end{eqnarray}
and
\begin{eqnarray}
\tau(u)
&=&
  -\frac{3\gamma\psi'(1+u)+\psi(1+u)[\frac{\pi^2}{6}+2\psi'(1+u)]}{u}
     \nonumber             \\
&&
  -\frac{[\psi(1+u)+\gamma]^2}{u^2}
  \nonumber \\
&&
  +\frac{\pi^2}{6u} \psi(1+u)
  +\int_0^1 \! \mbox{d}v \, \Xi(u,v),
\end{eqnarray}
with Lerch's transcendent $\phi(z,u,v)$ \cite{Magnus66} and
\begin{eqnarray}
 \Xi(&u,v&)
     \nonumber             \\
&= \!&
 \frac{\psi'(1+u)}{u \ln(v)}+
 \frac{1}{(1-v)u}
 \Bigg\{
  2v\phi(v,2,1+u)
     \nonumber             \\
&&
  +
  \frac{1-2v^u}{v}
  \left[ \ln(v)\phi(\frac{1}{v},1,1+u)+\phi(\frac{1}{v},2,1+u)\right]
     \nonumber    \\
&&
  +v^u\left[2\ln(v)\ln(1-v)
  +{1 \over 2} \ln^2(v)+3\mbox{Li}_2(1-v)\right]
      \nonumber             \\
&&
  -\frac{2(1-v^u)}{u}
  \left[\ln(1-v)+v\phi(v,1,1+u)\right]   
\Bigg\}.
\end{eqnarray}
The high temperature expansion of $Z$ is straightforward and reads
\begin{eqnarray}
 Z_{\rm SET}
 &=&  
 1 + g \sigma(u) (\beta E_C/2\pi^2)^2  +{\cal O}(\beta E_C)^3
     \nonumber             \\
 &&
  + 2 C_1 \cos(2 \pi n_g) \left[ 1+{\cal O}(\beta E_C)^2 \right],
\label{partition}
\end{eqnarray}
which combines with Eq.\ $(\ref{conductance})$ to yield an analytical
expression for the high temperature conduction of a SET valid for
arbitrary tunneling conductance.

\subsection{Discussion of Results and Comparison with Experimental Data}
\begin{figure}
\begin{center}
\leavevmode
\epsfxsize=0.5 \textwidth
\epsfbox{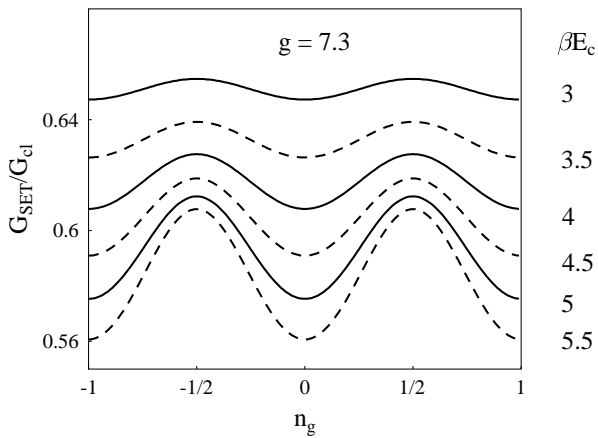}
\end{center}
\caption{Conductance of a symmetrical SET in dependence on the
dimensionless gate voltage $n_g$ for various temperatures and parallel
conductance $g=7.3$.}
\label{fig:fig13}
\end{figure}

In Fig.~\ref{fig:fig13} the normalized conductance 
$G_{\rm SET}/G_{\rm cl}$ is
depicted in dependence on the dimensionless gate voltage
$n_g$ for various temperatures $\beta E_C$. 
The quantum corrections are more pronounced for lower temperatures 
where the gate voltage dependence becomes more significant. The
oscillatory behavior of the conductance may be characterized in terms
of a maximum
$G_{\rm max}=G_{\rm SET}|_{n_g=1/2}$
and minimum 
$G_{\rm min}=G_{\rm SET}|_{n_g=0}$ linear conductance.

\begin{figure}
\begin{center}
\leavevmode
\epsfxsize=0.45 \textwidth
\epsfbox{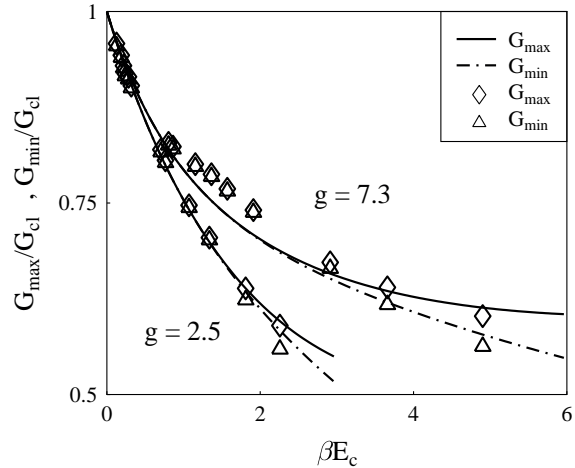}
\end{center}
\caption{Maximum and minimum linear conductance in dependence
on dimensionless temperature for two dimensionless parallel
conductances $g=2.5$ and $7.3$ compared with experimental data
(symbols) by Joyez {\it et al}. \protect\cite{JoyezSETPRL97}.}
\label{fig:fig14}
\end{figure}

\noindent
We have compared our findings for the maximum and minimum as a
function of temperature with recent experimental data by
Joyez et al.\ \cite{JoyezSETPRL97} 
for transistors with $g=0.6$, $2.5$ and $7.3$.
As seen from Fig.~\ref{fig:fig14} the theory describes the high
temperature behavior of all junctions (results for $g=0.6$ are not
shown) down to temperatures where the current starts to modulate
with the gate voltage. The parameters have {\it not} been 
adjusted to improve
the fit but coincide with the values given in \cite{JoyezSETPRL97}. 
The small
deviations between theory and experiment for $g=7.3$ near $\beta
E_C=1$ may arise from experimental uncertainties in $\beta E_C$
\cite{JoyezPriv}.
We mention
that the temperature dependence of the conductance of the highly
conducting SET ($g=7.3$) is not within reach of
previous theoretical work. The results obtained
should be useful for experimental studies
of even larger tunneling conductances since the predictions remain valid
for arbitrary values of $g$.

In the region of weak tunneling, $g<1$, the quantity $u$ becomes
small at high temperatures and we
may replace $\sigma(u)$ and $\tau(u)$ by
\begin{equation}
\sigma(0)=6 \zeta(3), \qquad
\tau(0)= \pi^4/10.
\end{equation}
This gives for the conductance
of a weakly conducting SET
\begin{eqnarray}
 \frac{G_{\rm SET}}{G_{\rm cl}}
 &=&
 \bigg[
    1-\frac{\beta E_C}{3}+\left(\frac{1}{15}
    +g \frac{3 \zeta(3)}{2\pi^4}\right) 
     \nonumber             \\
 &&   \times
  (\beta E_C)^2
    +{\cal O}(\beta E_C)^3
 \bigg],
\end{eqnarray}
in accordance with earlier work 
\cite{GeorgDipl96,JoyezHightempPRB97,KoenigZaikinPRB97}.
In the region of strong tunneling, the quantity $u$ is typically large
even for the highest temperatures explored experimentally and the full
expression $(\ref{conductance})$, $(\ref{partition})$ must be used.

\section{Conclusions}
In this article we have studied the conductance of 
nanofabricated metallic circuits
showing Coulomb blockade phenomena. We have treated electron
tunneling nonperturbatively based on a path integral expression 
derived in Sec.\ II. Then, the frequency dependent linear 
conductance of a single
tunnel junction embedded in an electromagnetic environment 
was calculated in the semiclassical approximation. We have shown that
this approximation is not only adequate for high temperatures but also
in the limit of large conductance.
As far as the leading quantum corrections are concerned, the tunnel
junction was shown to be described as an effective linear element with
an admittance that depends on the whole circuit. The predictions for
the dc conductance were compared with recent
experimental findings by two groups 
\cite{JoyezSJPRL98,ToppariSJEPL98} and we found
good agreement in the semiclassical regime of large conductance and/or
high temperatures. Further, we have shown that the low frequency
behavior of the ac conductance can be calculated in terms of a 
renormalized capacitance which shows a linear dependence on the 
inverse temperature. 

In Sec.\ IV we applied the method to a linear array of $N$ 
tunnel junctions and
determined the effect of the environmental impedance on the
conductance as well as the influence of
the array length $N$. For large $N$ the
conductance dip becomes independent of the electromagnetic environment in
accordance with previous calculations \cite{PekolaArrayJLTP97}.
For multi-junction circuits the configuration space of the phase
variables was shown to be a torus, and contributions of nonvanishing
winding numbers become relevant if one goes beyond the leading order
quantum corrections. 

The conductance of the single electron 
transistor was determined in Sec.\ V by including nontrivial 
winding numbers leading to the gate voltage dependence of the
conductance. 
The results were found to explain recent experimental data 
\cite{JoyezSETPRL97} for moderately low temperatures.
For lower temperatures the contribution of sluggon trajectories 
\cite{WangBOXPRB96} has to 
be taken into account which was not elaborated here. 

The semiclassical
theory presented has features in common with the
quasiclassical Langevin equation put forward in 
Ref.\ \cite{ZaikinSETJETP96}. 
The Gaussian approximation underlying
this approach is consistent with the semiclassical theory up to first 
order in $\beta E_C$. Our 
results consistently include higher order terms in $\beta E_C$ for
arbitrary tunneling conductances. Non-Gaussian fluctuations are
particularly relevant in the 
moderately large tunneling regime. 
The analytical
theory presented covers one edge in the temperature/conductance plane,
arbitrary
conductance and sufficiently high temperatures. 
Another edge is described by the
perturbative approach, arbitrary temperature and sufficiently small 
tunneling conductance. Both theories arise naturally from the formally
exact representation of the current-current correlator
which may also serve as a basis for 
Monte-Carlo simulations \cite{GeorgMCSVD99} 
that bridge between the semiclassical and
perturbative results. 

\section*{Acknowledgments}

The authors would like to thank Michel Devo\-ret, Daniel Esteve,
Shadyar Farhangfar, Philippe Joyez, Jukka Pekola, and Jussi Toppari 
for valuable discussions.
One of us (GG) acknowledges the hospitality of the CEA-Saclay
during an extended stay.
Financial support was provided by the Deutsche
Forschungsgemeinschaft (DFG) and the Deutscher Akademischer
Austauschdienst (DAAD).


\begin{references}
%
\bibitem{Nato92}
{\it Single Charge Tunneling}, edited by H.\ Grabert and M.H.\
Devoret, NATO ASI Series B, Vol.\ 294 (Plenum, New York, 1992).
%
\bibitem{FultonPRL87}
T.A.\ Fulton and G.J.\ Dolan,
Phys.\  Rev.\  Lett.\ {\bf 59}, 109 (1987).
%
\bibitem{DelsingFrequPRL89}
P.\ Delsing, K.K.\ Likharev, L.S.\ Kuzmin, and T.\ Claeson,
Phys.\ Rev.\ Lett.\ {\bf 63}, 1180 (1989).
%
\bibitem{KastnerRMP92}
 M.A.\ Kastner, Rev.\ Mod.\ Phys.\ {\bf 64}, 849 (1992).
%
\bibitem{PekolaSJPRL96}
J.P.\ Kauppinen and J.P.\ Pekola, 
Phys.\ Rev.\ Lett.\ {\bf 77}, 3889 (1996).
%
\bibitem{GrabertEMPRL90}
M.H.\ Devoret, D.\ Esteve, H.\ Grabert, G.-L.\ Ingold,
H.\ Pothier, and C.\ Urbina,
Phys.\ Rev.\ Lett.\ {\bf 64}, 1824 (1990).
%
\bibitem{Averin91}
D.V.\ Averin and K.K.\ Likharev, in {\it Mesoscopic Phenomena
in Solids}, ed.\ by B.L.\ Altshuler, P.A.\ Lee, and
R.A.\ Webb, p.\ 173 (North-Holland, Amsterdam, 1991).
%
\bibitem{SchoenEMPRB91}
A.A.\ Odintsov, G.\ Falci, and G. Sch\"on, 
Phys.\ Rev.\ B {\bf 44}, 13089 (1991).
%
\bibitem{IngoldNato92}
G.-L.\ Ingold and Yu.V.\ Nazarov, in Ref.\ 1.
%
\bibitem{NazarovTravPRB91}
Yu.V.\ Nazarov,
Phys.\ Rev.\ B {\bf 43}, R6220 (1991).
%
\bibitem{LafargeZPB91}
P.\ Lafarge, H.\ Pothier, E.R.\ Williams, D.\ Esteve,
C.\ Urbina, and M.H.\ Devoret,
Z.\ Phys.\ B {\bf 85}, 327 (1991).
%
\bibitem{EsteveDetAPL92}
A.N.\ Cleland, D.\ Esteve, C.\ Urbina, and M.H.\ Devoret,
Appl.\ Phys.\ Lett.\ {\bf 61}, 2820 (1992).
%
\bibitem{PekolaThermoPRL94}
J.P.\ Pekola, K.P.\ Hirvi, J.P.\ Kauppinen, and M.A.\ Paalanen, 
Phys.\ Rev.\ Lett.\ {\bf 73}, 2903 (1994).
%
\bibitem{AverinNato92}
D.V.\ Averin and Yu.V.\ Nazarov, in Ref.\ 1.
%
\bibitem{KoenigCOTPRL97}
J.\ K\"onig, H.\ Schoeller, and G.\ Sch\"on,
Phys.\ Rev.\ Lett.\ {\bf 78}, 4482 (1997).
%
\bibitem{JoyezSETPRL97}
P.\ Joyez, V.\ Bouchiat, D.\ Esteve, C.\ Urbina, 
and M.H.\ Devoret, 
Phys.\ Rev.\ Lett.\ {\bf 79}, 1349 (1997).
%
\bibitem{JoyezSJPRL98}
P.\ Joyez, D.\ Esteve, and M.H.\ Devoret, 
Phys.\ Rev.\ Lett.\ {\bf 80}, 1956 (1998).
%
\bibitem{ToppariSJEPL98}
Sh.\ Farhangfar, J.J.\ Toppari, Yu.A.\ Pashkin, A.J.\ Manninen,
and J.P.\ Pekola, Europhys.\ Lett.\ {\bf 43}, 59 (1998).
%
\bibitem{KuzminSETPRB99}
D.\ Chouvaev, L.S.\ Kuzmin, D.S.\ Golubev, and A.D.\ Zaikin,
Phys.\ Rev.\ B {\bf 59}, 10599 (1999).
%
\bibitem{GrabertBOXPRB94}
H.\ Grabert, Phys.\ Rev.\ B {\bf 50}, 17364 (1994).
%
\bibitem{GrabertBOXPhysica94}
H.\ Grabert, Physica {\bf B 194-196}, 1011 (1994).
%
\bibitem{GeorgBOXPRL98}
G.\ G\"oppert, H.\ Grabert, N.V.\ Prokof'ev, and
B.V.\ Svistunov, 
Phys.\ Rev.\ Lett.\ {\bf 81}, 2324 (1998).
%
\bibitem{KoenigRENPRB98}
J.\ K\"onig, H.\ Schoeller, and G.\ Sch\"on,
Phys.\ Rev.\ B {\bf 58}, 7882 (1998).
%
\bibitem{GeorgChannelEPL99}
G.\ G\"oppert, H.\ Grabert, and C. Beck,
Europhys.\ Lett.\ {\bf 45}, 249 (1999).
%
\bibitem{MatveevBOXJETP91}
K.A.\ Matveev, 
Sov.\ Phys.\ JETP\ {\bf 72}, 892 (1991);
%
\bibitem{Schoeller2StatePRB94}
H.\ Schoeller and G. Sch\"on,
Phys.\ Rev.\ B {\bf 50}, 18436 (1994).
%
\bibitem{FalciScalePRL95}
G.\ Falci, G.\ Sch\"on, and G.T.\ Zimanyi,
Phys.\  Rev.\  Lett.\ {\bf 74}, 3257 (1995).
%
\bibitem{ZaikinNCAPRB94}
D.S.\ Golubev and A.D.\ Zaikin, 
Phys.\ Rev.\ B {\bf 50}, 8736 (1994).
%
\bibitem{KoenigRGPRL98} 
J.\ K\"onig and H.\ Schoeller, 
Phys.\ Rev.\ Lett.\ {\bf 81}, 3511 (1998).
%
\bibitem{PohjolaDoublePRB99}
T.\ Pohjola, J.\ K\"onig, H.\ Schoeller, and G.\ Sch\"on,
Phys.\ Rev.\ B {\bf 59}, 7579 (1999).
%
\bibitem{SchoenREP90}
G.\ Sch\"on and A.D.\ Zaikin, Phys.\ Rep.\ {\bf 198}, 237 (1990).
%
\bibitem{BenJacPRL83}
E.\ Ben-Jocab, E.\ Mottola, and G.\ Sch\"on,
Phys.\  Rev.\  Lett.\ {\bf 51}, 2064 (1983).
%
\bibitem{ZaikinECPRL91}
S.V.\ Panyukov and A.D.\ Zaikin,
Phys.\ Rev.\ Lett.\ {\bf 67}, 3168 (1991).
%
\bibitem{ZaikinSJPRB92}
D.S.\ Golubev and A.D.\ Zaikin, 
Phys.\ Rev.\ B {\bf 46}, 10903 (1992).
%
\bibitem{ZaikinSETJETP96}
D.S.\ Golubev and A.D.\ Zaikin, 
JETP Lett. {\bf 63}, 1007 (1996).
%
\bibitem{WangBOXPRB96}
X.\ Wang and H.\ Grabert,
Phys.\ Rev.\ B {\bf 53}, 12621 (1996).
%
\bibitem{WangMCEPL97}
X.\ Wang, R.\ Egger, and H.\ Grabert,
Europhys.\ Lett.\ {\bf 38}, 545 (1997).
%
\bibitem{ZwergerBOXPRL97}
W.\ Hofstetter and W.\ Zwerger, 
Phys.\ Rev.\ Lett.\ {\bf 78}, 3737 (1997).
%
\bibitem{HerreroMCPRB99}
C.P.\ Herrero, G.\ Sch\"on, and A.D.\ Zaikin,
Phys.\ Rev.\ B {\bf 59}, 5728 (1999).
%
%
%
%
\bibitem{GeorgSJPRB97}
G.\ G\"oppert, X.\ Wang, and H.\ Grabert,
Phys.\ Rev.\ B {\bf 55}, R10213 (1997).
%
\bibitem{GeorgSETPRB98}
G.\ G\"oppert and H.\ Grabert,
Phys.\ Rev.\ B {\bf 58}, R10155 (1998).
%
\bibitem{GeorgSEMICRAS99}
G.\ G\"oppert and H.\ Grabert,
C.\ R.\ Acad.\ Sci.\ {\bf 327}, 885 (1999).
%
\bibitem{DolanLithoPhysica88}
G.J.\ Dolan and J.H.\ Dunsmuir,
Physica B {\bf 152}, 7 (1988).
%
\bibitem{BardeenTunnelPRL61}
J.\ Bardeen,
Phys.\ Rev.\ Lett.\ {\bf 6}, 57 (1961).
%
\bibitem{CaldeiraAP83}
A.O.\ Caldeira and A.J.\ Leggett, 
Ann.\ Phys.\ (N.Y.) 149, 374 (1983).
%
\bibitem{BaymGreensJMP61}
G.\ Baym and N.D.\ Mermin,
J.\ Math.\ Phys.\ {\bf 2}, 232 (1961).
%
\bibitem{GrabertREP88}
H.\ Grabert, P.\ Schramm, and G.-L.\ Ingold,
Phys.\ Rep.\ {\bf 168}, 115 (1988).
%
\bibitem{GeorgSCHA99}
G.\ G\"oppert and H.\ Grabert, (to be published).
%
\bibitem{PekolaArrayJLTP97}
Sh.\ Farhangfar, K.P.\ Hirvi, J.P.\ Kauppinen, J.P.\ Pekola,
J.J.\ Toppari, D.V.\ Averin, and A.N.\ Korotkov,
J.\ Low Temp.\ Phys.\, {\bf 108}, 191 (1997). 
%
%
\bibitem{Farhangfar99}
Sh.\ Farhangfar, A.J.\ Manninen, and J.P.\ Pekola,
cond-mat/9910238.
%
\bibitem{GeorgMCSVD99}
G.\ G\"oppert, B.\ H\"upper, and H.\ Grabert,
(to be published).
%
\bibitem{Abramowitz70}
M.\ Abramowitz, I.A.\ Stegun,
{\it Handbook of Mathematical Functions}
(Dover, New York, 1970).
%
\bibitem{Magnus66}
W.\ Magnus, F.\ Oberhettinger, and R.P.\ Soni,
{\it Formulas and Theorems for the Special Functions of Mathematical
Physics} (Springer, Berlin, 1966)
%
\bibitem{JoyezPriv}
P.\ Joyez, private communication.
%
\bibitem{GeorgDipl96}
G.\ G\"oppert, Diploma thesis, Freiburg 1996.
%
\bibitem{JoyezHightempPRB97}
P.\ Joyez and D.\ Esteve, 
Phys.\ Rev.\ B {\bf 56}, 1848 (1997).
%
\bibitem{KoenigZaikinPRB97}
D.S.\ Golubev, J.\ K\"onig, H.\ Schoeller, G.\ Sch\"on, and
A.D.\ Zaikin, 
Phys.\ Rev.\ B {\bf 56}, 15782 (1997).
%


\end{references}


\end{multicols}
\end{document}